\documentclass[fleqn,usenatbib]{mnras}

\usepackage{newtxtext,newtxmath}

\usepackage[T1]{fontenc}
\usepackage{ae,aecompl}

\usepackage{graphicx}	
\usepackage{amsmath}	
\usepackage{amssymb}	
\usepackage{float}
\usepackage{rotating}
\usepackage{subcaption}

\usepackage{etoolbox}
\makeatletter
\patchcmd\@combinedblfloats{\box\@outputbox}{\unvbox\@outputbox}{}{%
 \errmessage{\noexpand\@combinedblfloats could not be patched}%
}%
\makeatother

\newcommand{\cha}{{\it Chandra}}
\newcommand{\Msun}{${\rm M}_{\odot}$}

\newcommand\phn{\phantom{0}}%
\title[{\em Swift} Bulge Survey: Opt/NIR follow-up]{The {\em Swift} Bulge Survey: optical and near-IR follow-up featuring a likely symbiotic X-ray binary \& a focused wind CV}

\author[A. W. Shaw et al.]{%
A. W. Shaw,$^{1,2}$\thanks{E-mail: aarrans@unr.edu}
C. O. Heinke,$^{1}$
T. J. Maccarone,$^{3}$
G. R. Sivakoff,$^{1}$
J. Strader,$^{4}$
\newauthor A. Bahramian,$^{5}$
N. Degenaar,$^{6}$
J. A. Kennea,$^{7}$
E. Kuulkers,$^{8}$
A. Rau,$^{9}$
\newauthor L. E. Rivera Sandoval,$^{3}$
L. Shishkovsky,$^{4}$
S. J. Swihart,$^{4}$
A. J. Tetarenko,$^{10}$
\newauthor R. Wijnands,$^{6}$
and J. J. M. in 't Zand$^{11}$
\\
$^{1}$Department of Physics, University of Alberta, CCIS 4-181, Edmonton, AB T6G 2E1, Canada\\
$^{2}$Department of Physics, University of Nevada, Reno, NV 89557, USA\\
$^{3}$Department of Physics, Box 41051, Science Building, Texas Tech University, Lubbock, TX 79409-1051, USA\\
$^{4}$Center for Data Intensive and Time Domain Astronomy, Department of Physics and Astronomy, Michigan State University,\\ East Lansing, MI 48824, USA\\
$^{5}$International Centre for Radio Astronomy Research-Curtin University, GPO Box U1987, Perth, WA 6845, Australia\\
$^{6}$Anton Pannekoek Institute for Astronomy, University of Amsterdam, Postbus 94249, NL-1090 GE Amsterdam, the Netherlands\\
$^{7}$Department of Astronomy and Astrophysics, The Pennsylvania State University, University Park, PA 16802, USA\\
$^{8}$ESA/ESTEC, Keplerlaan 1, 2201, AZ Noordwijk, the Netherlands\\
$^{9}$Max-Planck Institute for Extraterrestrial Physics, Giessenbachstr. 1, D-85748 Garching, Germany\\
$^{10}$East Asian Observatory, 660 N. A'oh\={o}k\={u} Place, University Park, Hilo HI, USA, 96720\\
$^{11}$SRON Netherlands Institute for Space Research, Sorbonnelaan 2, NL-3584 CA Utrecht, the Netherlands
}

\date{Accepted XXX. Received YYY; in original form ZZZ}

\pubyear{2019}

\begin{document}
\label{firstpage}
\pagerange{\pageref{firstpage}--\pageref{lastpage}}
\maketitle

\begin{abstract}
The nature of very faint X-ray transients (VFXTs) -- transient X-ray sources that peak at luminosities $L_X\lesssim10^{36} {\rm \, erg \, s^{-1}}$ -- is poorly understood. The faint and often short-lived outbursts make characterising VFXTs and their multi-wavelength counterparts difficult. In 2017 April we initiated the {\em Swift} Bulge Survey, a shallow X-ray survey of $\sim$16 square degrees around the Galactic centre with the {\em Neil Gehrels Swift Observatory}. The survey has been designed to detect new and known VFXTs, with follow-up programmes arranged to study their multi-wavelength counterparts. Here we detail the optical and near-infrared follow-up of four sources detected in the first year of the Swift Bulge Survey. 
The known neutron star binary IGR J17445-2747 has a K4III donor, indicating a potential symbiotic X-ray binary nature and the first such source to show X-ray bursts. We also find one nearby M-dwarf (1SXPS J174215.0-291453) and one system without a clear near-IR counterpart (Swift J175233.9-290952). Finally, 3XMM J174417.2-293944 has a subgiant donor, an 8.7 d orbital period, and a likely white dwarf accretor; we argue that this is the first detection of a white dwarf accreting from a gravitationally focused wind.
A key finding of our follow-up campaign is that binaries containing (sub)giant stars may make a substantial contribution to the VFXT population.
\end{abstract}

\begin{keywords}
X-rays: binaries -- infrared: stars -- surveys -- stars: neutron -- binaries: symbiotic -- novae, cataclysmic variables
\end{keywords}



\section{Introduction}

The majority of the brightest Galactic X-ray transients are binary systems in which a compact object, either a black hole (BH) or a neutron star (NS), accretes matter from a stellar companion. 
Large variations in mass accretion rate cause their X-ray luminosities to increase by factors of $>10^3$ and exceed $L_X \gtrsim 10^{36} {\rm \, erg \, s^{-1}}$ during outburst episodes, with a similar response at optical, near-infrared (NIR) and ultraviolet (UV) wavelengths \citep[see e.g.][]{Kuulkers-1998,Zurita-2006,Tucker-2018}. However, a fainter class of X-ray  transients, very faint X-ray transients (VFXTs), display lower peak X-ray luminosities of $\sim10^{34-36} {\rm \, erg \, s^{-1}}$ and corresponding faint optical/NIR outbursts \citep[e.g.][]{int-Zand-1999,Cornelisse-2002a,int-Zand-2005a,Wijnands-2006,Degenaar-2009,Armas-Padilla-2013a,Shaw-2018}. Whilst classical X-ray transients are relatively well studied, the discovery and study of VFXTs has been hampered by sensitivity limitations of the X-ray, optical and NIR instruments that are best-suited to time domain work.  Consequently the nature of VFXTs, and the cause of their sub-luminous outbursts, are not well understood.

Suggestions for the nature of VFXTs fall broadly into two groups: those with BH or NS accretors in some unusual state and other systems, often accreting white dwarf (WD) systems, such as intermediate polars \citep[many of which reach $L_X>10^{34} {\rm \, erg \, s^{-1}}$ and tend to be persistent, though some exhibit dwarf nova outbursts;][]{Mukai-2017,Schwope-2018}, novae \citep{Mukai-2008}, and symbiotic stars \citep{Luna-2013,Mukai-2016,Yungelson-2019}. 

Explaining VFXTs as (NS or BH) X-ray binaries with unusual properties requires that the accretion flow, or its interaction with the compact object, be modified. 
The outburst-quiescence cycle of many disc-accreting compact objects can broadly be described by the disc instability model \citep[DIM; see e.g.][for a review]{Lasota-2001}. In this framework, accumulation of matter leads to the accretion disc reaching a critical temperature (the ionization temperature of the dominant species in the accretion disc), triggering a bright outburst due to the increased strength of the magneto-rotational instability in ionized gas relative to neutral gas, after which the disc cools as the source returns to quiescence. The DIM has been shown to reproduce the global behaviour of a number of transient and persistent low-mass X-ray binaries \citep[LMXBs; e.g.][]{Coriat-2012,TetarenkoB-2016}. 
However, it is difficult for the DIM to explain the faintness of VFXT outbursts, and their low time-averaged mass transfer rate, if they are produced by X-ray binaries similar to those that have been well-studied \citep{Hameury-2016}. 

The faintness of transient VFXT outbursts implies very small accretion discs, suggestive of short (e.g.\ $<$2 h) orbital periods \citep{Shahbaz-1998c,King-2006,Heinke-2015}, interference with the flow of accreting matter onto the accretor, or that only a small part of a larger accretion disc is drained \citep{Degenaar-2009}. 

The low time-averaged inferred mass transfer rates of VFXTs have inspired a range of possible  explanations. It is possible that the true mass transfer rates from the donor are higher than we infer, implying accretion is hidden or that matter is thrown out before accreting onto the primary. 
Some VFXTs are known to be seen at high inclination \citep[such that we are only viewing scattered X-rays; e.g.][]{Muno-2005b,Corral-Santana-2013}, but statistical arguments rule out this being the predominant VFXT mechanism \citep{Wijnands-2006}.
BH LMXBs at low mass transfer rates may be particularly faint, if their radiative efficiency continues to decline with luminosity \citep{Maccarone-2013,Knevitt-2014}.  Alternatively, mass transfer could be highly non-conservative \citep[e.g.][]{Hernandez-Santisteban-2019}.  
One scenario for quasi-persistent VFXTs  \citep[e.g.][]{Heinke-2009,Heinke-2015} is that a strong magnetic field could impede the flow of matter onto the primary through a process known as the propeller effect \citep{Illarionov-1975}. A rapidly rotating magnetic field in this case removes a large amount of the mass transferred from the companion, whilst some material may be able to reach the NS poles \citep{Romanova-2005}, resulting in an overall low accretion rate. The identification of three ``transitional" (switching between X-ray active and radio pulsar states) millisecond pulsars, which appear to accrete\footnote{\citep[or possibly stimulate enhanced radio pulsar activity; see][]{Ambrosino-2017}} quasi-persistently at $L_X\sim10^{33-34} {\rm \, erg \, s^{-1}}$ \citep{Papitto-2013,Bogdanov-2015,deMartino-2013} shows that this is a plausible theory for other VFXTs.

Alternatively, the observed low mass transfer rates could reflect the real mass transfer rates from the donor. \citet{King-2006} suggested that mass transfer rates in VFXTs were so low as to require truly extreme scenarios, such as 100-1000 \Msun\ BH accretors. Several authors have suggested accretion from the wind of a main-sequence (or subgiant) donor star \citep{Bleach-2002,Pfahl-2002,Maccarone-2013}. However, \citet{Heinke-2015} showed that the low  mass transfer rates seen in the Galactic Centre VFXTs can be explained by standard binary evolution involving either normal H-rich donors,  or H-poor (white dwarf) donors, in either case culminating in $\ll0.1$ \Msun, partly degenerate donor stars in $\sim$2 h, or $\sim$90 min (respectively), orbits.  On the other hand, it is clear that these short-orbit systems cannot make up most VFXTs, based on evidence from several VFXT optical/infrared counterparts \citep{Degenaar-2010a,Shaw-2017c}.

To develop a complete understanding of VFXTs we  require dedicated multi-wavelength studies of a substantial population. In particular, the companion stars of X-ray binaries are usually only accessible through optical/NIR follow-up of sources discovered in X-ray surveys. Furthermore, it is well-established that VFXTs represent a heterogeneous class of systems, and that an extensive survey is necessary to unveil the relative proportions of different mechanisms.

\section{The Swift Bulge Survey}
\label{sec:SBS}

The majority of VFXTs so far discovered have been found through frequent monitoring of the dense region around the Galactic Centre \citep{Muno-2005a,Wijnands-2006,Degenaar-2009,Degenaar-2010b,Degenaar-2012c}. 
However, the Galactic Centre is inaccessible to optical and infrared follow-up of typical X-ray binary companions, due to its extreme extinction and stellar crowding \citep{Bandyopadhyay-2005,Laycock-2005,DeWitt-2010}.
We therefore planned a survey designed to repeatedly monitor a large swath of the Galactic Bulge, much of which experiences low enough reddening and stellar crowding to enable plausible optical/infrared follow-up.
To this end, in 2017 April we initiated the {\em Swift} Bulge Survey \citep[SBS; see e.g.][]{Heinke-2017,Bahramian-2017b}, a shallow (60 s exposures) monitoring campaign of $\sim16$ square degrees around the Galactic Centre with the {\em Neil Gehrels Swift Observatory}/X-ray Telescope and Ultraviolet/Optical Telescope \citep[{\em Swift}/XRT and {\em Swift}/UVOT;][]{Burrows-2005,Roming-2005}, designed to detect new and known VFXTs and study X-ray variability in the direction of the Galactic Bulge in the low flux regime \citep[see][in prep., for details]{Bahramian-2020-prep}. In the first year of this survey 2017 April -- 2018 March), we obtained a total of 19 biweekly epochs. In addition to the SBS observations, we also obtained target of opportunity (ToO) triggers with a multitude of optical, NIR, radio and X-ray facilities to follow up sources detected in the X-ray survey in an effort to derive the natures of such sources. The details of the X-ray survey are described by \citet[][in prep.]{Bahramian-2020-prep} and the UV results are presented by \citet[][in prep.]{Rivera-Sandoval-2020-prep}. In this work we describe the results from our optical/NIR follow-up campaign, the targets of which are as follows: previously known sources {\bf IGR\,J17445$-$2747}, {\bf 1SXPS\,J174215.0$-$291453} and {\bf 3XMM\,J174417.2$-$293944} and the newly discovered source {\bf Swift\,J175233.9$-$290952}. The targets are described in more detail in Appendix \ref{app:sources} \citep[see also][in prep.]{Bahramian-2020-prep}.

In Section \ref{sec:Obs} we describe the follow-up observations we have utilized in this work. 
In Section \ref{sec:Analysis} we present our analysis techniques and the results 
from the observational data. We discuss the implications 
in Section \ref{sec:Discussion} and provide a 
summary in Section \ref{sec:Conclusions}.

\section{Observations and Data Reduction}
\label{sec:Obs}


We used several optical and NIR facilities to follow up X-ray sources detected in the SBS.We detail here the follow-up observations and the steps we took to reduce and calibrate the data. We summarise our follow-up observations in Table \ref{tab:follow-up}.

\begin{table*}
    \centering
    \caption{Summary of optical/NIR observations of sources detected in the SBS that we utilized in this work. For each X-ray source we also give the coordinates, positional uncertainty and the name of the X-ray observatory that constrained the position.} 
    \begin{tabular}{cccc}
    \hline
    \hline
        \multicolumn{4}{c}{{\bf IGR\,J17445$-$2747} \qquad $17^{\rm h}44^{\rm m}30\fs437$ $-27^\circ46'00\farcs32$ (1$\arcsec$; \cha)}\\
    \hline
         Date & Telescope/Instrument & Filter/Grating & Exposure Time  \\
    \hline
        2017 April 18 & {\em Gemini}/NIRI & $J$, $H$, $K_s$ & 275 s, 275 s, 275 s\\
        2017 April 24 & {\em SOAR}/Goodman & 400 l mm$^{-1}$ & $1800$ s\\
        2018 May 09 & {\em Gemini}/Flamingos-2 & R3K $J$, $K_s$ & 1200 s, 600 s \\
    \hline
    \hline
        \multicolumn{4}{c}{{\bf Swift\,J175233.9$\mathbf-$290952} \qquad $17^{\rm h}52^{\rm m}33\fs934$ $-29^\circ09'47\farcs92$ (0\farcs7; \cha)}\\
    \hline
         Date & Telescope/Instrument & Filter/Grating & Exposure Time  \\
    \hline
        2017 May 9 & MPI/ESO 2.2m/GROND & $J$, $H$, $K$ & 560 s, 480 s, 560 s \\
        2017 June 26 & {\em VLT}/SINFONI & $H+K$ & 4560 s \\ 
    \hline
    \hline
        \multicolumn{4}{c}{{\bf 1SXPS\,J174215.0$\mathbf-$291453} \qquad $17^{\rm h}42^{\rm m}14\fs995$ $-29^\circ14'59\farcs40$ (0\farcs4; {\em XMM-Newton})}\\
    \hline
         Date & Telescope/Instrument & Filter/Grating & Exposure Time  \\
    \hline
        2016 March 10 -- 2018 September 23 & ASAS-SN & $V$ & 90 s (per exposure)\\
        2017 May 28 & {\em SOAR}/Goodman & 400 l mm$^{-1}$ & 300 s \\
    \hline
    \hline
        \multicolumn{4}{c}{{\bf 3XMM\,J174417.2$\mathbf-$293944} \qquad $17^{\rm h}44^{\rm m}17\fs246$ $-29^\circ39'44\farcs31$ (0\farcs5; {\em XMM-Newton})}\\
    \hline
         Date & Telescope/Instrument & Filter/Grating & Exposure Time  \\
    \hline
         2016 March 10 -- 2018 September 23 & ASAS-SN & $V$ & 90 s (per exposure)\\
         2017 September 29 & {\em VLT}/SINFONI & $H+K$ & 1280 s \\
         2018 March 25 -- September 15 & {\em SOAR}/Goodman & 2100 l mm$^{-1}$ & 600 s (per night)$^{a}$ \\
    \hline
    \hline
    \end{tabular}
    \\
    \small$^{a}$ Typical setup, see text for further details and exceptions
    \label{tab:follow-up}
\end{table*}

\subsection{NIR photometry}
\label{sec:nir_im}
We obtained NIR images of the IGR\,J17445$-$2747 field on 2017 April 18 with the Near InfraRed Imager and spectrograph (NIRI) on the 8.1m {\em Gemini} North telescope at Mauna Kea, Hawaii as part of proposal GN-2017A-Q-259 (PI: Bahramian). The instrument was operating in imaging mode with the f/6 camera. 
We obtained 10 exposures of 27.5 s in the $J$, $H$ and $K$ broad-band filters. To account for the changing sky background at NIR wavelengths, a dithering pattern was applied in each filter, with each co-added exposure consisting of 25 exposures of 1.1s. 

To reduce the data we used the Image Reduction and Analysis Facility \citep[{\sc iraf};][]{Tody-1986} {\em Gemini} package, in conjunction with NIRI-specific {\sc python} routines. We utilized the {\tt cleanir} script\footnote{\href{http://staff.gemini.edu/~astephens/niri/cleanir/cleanir.py}{http://staff.gemini.edu/\textasciitilde astephens/niri/cleanir/cleanir.py}} to remove artifacts superimposed by the IR detector controller, and corrected for non-linearity in the detector with {\tt nirlin}.\footnote{\href{http://staff.gemini.edu/\~astephens/niri/nirlin}{http://staff.gemini.edu/\textasciitilde astephens/niri/nirlin}} Sky frames were created from the science images, as there were no extended objects in the field. For each target, a normalized flat-field was created with the task {\tt niflat} and bad pixels identified using short dark frames. Flat-fielding and sky subtraction was performed using {\tt nireduce} and created final images  with {\tt imcoadd}.

To derive the correct astrometry  we first used {\sc SExtractor} \citep{Bertin-1996} to create source catalogues for the images in each band. The astrometric solution was calculated with {\sc scamp} \citep{Bertin-2006} using 2MASS as a reference, providing a typical astrometric accuracy of $0\farcs01$ in both RA and Dec. The astrometric solution was then mapped to the co-added images with {\sc swarp} \citep{Bertin-2002}.

We utilized the {\sc iraf daophot} routines developed for crowded field photometry \citep{Stetson-1987}. We determined the empirical point spread function (PSF) for each target frame using 15 relatively isolated field stars present in the UKIDSS catalogue. Using the fitted PSF, we subtracted close neighbors of the UKIDSS stars from the images, and used these (now isolated) stars to calibrate the photometry. We performed aperture photometry on both the target and the calibration stars with the task {\tt phot}, employing a 10 pixel ($1\farcs16$) radius circular aperture for the target and a 15 pixel ($1\farcs75$) radius for the calibration stars. The appropriate aperture correction was computed with the task {\tt mkapfile}. To convert between instrumental magnitudes and the UKIDSS magnitude system we used the {\sc iraf} task {\tt fitparams} and transformation equations of the form

\begin{equation}
	j = J + c_1 + c_2X_j + c_3(J-H),
\end{equation}
\begin{equation}
    h = H + c_4 + c_5X_h + c_6(H-K), {\rm and}
\end{equation}
\begin{equation}
    k = K + c_7 + c_8X_k + c_9(H-K),
\end{equation}

\noindent where, $j,h,k$ are the instrumental magnitudes in the appropriate filter, $J,H,K$ are the known magnitudes of the calibration stars, $c_{1-9}$ are constants (representing an additive term, an extinction term and a colour term in each band) and $X_{j,h,k}$ are the average airmasses for observations in each band. We estimate uncertainties on the measured magnitudes of the target by calculating the root-mean-square (RMS) error, comparing the catalogue magnitudes of the calibration stars with those derived by our photometry.

We obtained NIR imaging observations of the Swift\,J175233.9$-$290952 field on 2017 May 9 with the Gamma-Ray Burst Optical/Near-Infrared Detector \citep[GROND;][PI: Rau]{Greiner-2008a} mounted on the 2.2m MPI/ESO telescope at La Silla, Chile. We obtained simultaneous imaging in the $J$, $H$, and $K$-bands, consisting of 56, 10 s exposures in each of $J$ and $K$ and 48, 10 s exposures in $H$. Data were reduced and photometry was performed with the GROND manual analysis pipeline v2.2, which used 2MASS stars in the field to derive the astrometric solution, achieving an astrometric accuracy of $0\farcs03$ or better in each band. Instrumental magnitudes were calibrated on to the 2MASS photometric system using the known NIR transformations \citep{Greiner-2008a}.

\subsection{NIR spectroscopy}
\label{sec:nir_spec}
On 2017 June 26 we obtained spectroscopy of the region surrounding the \cha\ position of Swift\,J175233.9$-$290952 with the Spectrograph for INtegral Field Observations in the Near Infrared \citep[SINFONI;][]{Eisenhauer-2003} integral field unit (IFU) 
on the {\em Very Large Telescope} ({\em VLT}) at Paranal, Chile. Target of Opportunity (ToO) observations were obtained as part of proposal 099.D-0826 (PI: Degenaar). We obtained 19 images, each a co-add of 4 dithered 60 s exposures (to account for the variable NIR sky background), using the $H+K$ grating. We utilized the $0\farcs25$ pixel scale, which 
provides a typical full width at half maximum (FWHM) resolution of 11.8 \AA. 
We used a nearby $R=13.7$ star as a natural guide star (NGS) to correct for atmospheric distortions with the SINFONI adaptive optics (AO).

Data were reduced using the ESO Recipe Execution tool ({\sc EsoRex}) pipeline\footnote{\href{https://www.eso.org/sci/software/cpl/esorex.html}{https://www.eso.org/sci/software/cpl/esorex.html}}, which performs typical reduction steps including dark subtraction, non-linearity correction, flat fielding and wavelength calibration to provide a co-added data cube. The astrometric solution for the median image was derived with the {\sc iraf} task {\tt ccmap}, using 6 UKIDSS stars in the small FOV image. The solution provided an accuracy of $0\farcs03$ and $0\farcs01$ in RA and Dec, respectively. We extracted the spectrum of the suspected counterpart with the {\sc EsoRex} task {\tt sinfo\_utl\_cube2spectrum}, using a circular region with a 3 pixel ($0\farcs375$) radius.

To correct for (terrestrial) telluric features of the {\em VLT} spectra we used {\sc molecfit} \citep{Smette-2015,Kausch-2015}, which fits a number of telluric lines in the science spectrum to derive a telluric transmission spectrum for the entire spectral range. This spectrum is then divided through the science spectrum, resulting in a spectrum relatively free of telluric absorption. {\sc molecfit} is able to use the science spectrum as an input, rather than a telluric standard, meaning that the transmission spectrum we derive is representative of the NIR sky spectrum under the actual observing conditions, 
 as opposed to telluric standard stars observed at different times, airmasses, and sky conditions.
 
We obtained IFU spectroscopy of the 3XMM\,J174417.2$-$293944 field with VLT/SINFONI on 2017 September 29.
We obtained 32 images, each a co-add of 4 dithered 10 s exposures, using the H+K grating. We utilized the $0\farcs1$ spatial pixel scale, providing a 
typical FWHM resolution of 10 \AA.  We used the target itself as the NGS for the AO correction.
Data were reduced in the same manner as  for Swift\,J175233.9$-$290952,  using a circular region with a 6 pixel ($0\farcs3$) radius centred on the star. The spectrum was corrected for telluric features with {\sc molecfit}, ensuring that stellar lines were excluded from the fit.

On 2018 May 09 we obtained long slit spectroscopy of 2MASS\,J17443041$-$2746004, the suspected NIR counterpart to IGR\,J17445$-$2747 \citep{Shaw-2017b}, with Flamingos-2 on the 8.1m {\em Gemini} South telescope at Cerro P\'{a}chon, Chile, as part of  the poor weather proposal GS-2018A-Q-407 (PI: Shaw). We obtained 2x600 s exposures in the $J$-band and 2x300 s exposures in the $K_s$-band, utilising the R3K disperser in both instances, providing a typical FWHM resolution of 4 \AA\ in the $J$-band and 7 \AA\ in the $K_s$-band. The second exposure in each band was offset along the two pixel ($0\farcs36$) wide slit by $10\arcsec$ to aid sky subtraction by accounting for the changing sky background at NIR wavelengths. We also obtained 4x30 s observations of the telluric standard A0V star HD\,155379 in each band. 

Data were reduced using the {\sc iraf} {\em Gemini} package. Flat fielding and dark subtraction was performed with the task {\tt nsreduce}. The wavelength solution was obtained with {\tt nswavelength} and applied to all telluric and science frames before spectra were extracted with {\tt nsextract}. We normalised the averaged spectrum of the telluric standard and removed the hydrogen absorption features at Brackett $\gamma$ (Br$\gamma$; 2.166 $\mu$m) and Paschen $\beta$ (Pa$\beta$; 1.282 $\mu$m) with a best-fit Voigt profile. We then used the {\sc iraf} task {\tt telluric}, which shifts and scales the science and telluric spectra to best divide out telluric features from the science spectra. 
 
\subsection{Optical Spectroscopy}
\label{sec:optical_spec}
On 2017 April 24 we obtained a single 1800 s spectrum of {\em Gaia} DR2 4060626256817246720, the faint ($i\sim19.4$) optical counterpart consistent with the \cha\ position of IGR\,J17445$-$2747, with the Goodman Spectrograph \citep{Clemens-2004} on the 4.1m {\em SOuthern Astrophysical Research} ({\em SOAR}) telescope, on Cerro Pach\'{o}n, Chile (PI: Strader). We used a $0\farcs95$ slit and a 400 line mm$^{-1}$ grating, with an approximate wavelength coverage of $\sim 4800$--8800 \AA\ at a FWHM resolution of 5.6 \AA. 
The 1D spectrum was optimally extracted from the 2D CCD image using the {\sc iraf} task {\tt apall}. The target spectrum was wavelength calibrated using the task {\tt identify} and a single spectrum of the FeAr lamp. Flux calibration of the target spectrum was performed using the flux standard star CD-32 9927.

On 2017 May 28 we obtained a single 300 s spectrum of {\em Gaia} DR2 4057126472597377152, the optical counterpart to 1SXPS\,J174215.0$-$291453, with {\em SOAR}/Goodman. We used an identical setup to the {\em SOAR}/Goodman observations of IGR\,J17445$-$2747, and reduction was carried out in the same manner, including using the same flux standard star.

We obtained spectra of {\em Gaia} DR2 4057051396569058432, the optical counterpart to 3XMM\,J174417.2$-$293944, with {\em SOAR}/Goodman over 17 epochs from 2018 Mar 25 to 2018 Sep 15 UTC. At each epoch we obtained two  (typically 300 s) spectra back to back. 
Most observations used a 2100 line mm$^{-1}$ grating and a 0\farcs95 slit, yielding a resolution of 0.8 \AA\ FWHM  over 
$\sim 6030-6615$ \AA. On 2018 Aug 25, we instead used a narrower 0\farcs45 slit to yield an improved resolution of 0.5 \AA. 
On 2018 Aug 13 only a 1200 line mm$^{-1}$ grating was available, which gave a lower resolution of 1.7 \AA. All spectra were reduced and optimally extracted in the same manner as IGR\,J17445$-$2747 and 1SXPS\,J174215.0$-$291453. To compute radial velocities (RVs), we also observed a K giant with the same setup.

\subsection{Long-term Optical Monitoring}
\label{sec:optical_monitoring}
The optical counterparts to 1SXPS\,J174215.0$-$291453 
and 3XMM\,J174417.2$-$293944 
are monitored in the $V$-band by the All-Sky Automated Survey for Supernovae \citep[ASAS-SN;][]{Shappee-2014,Kochanek-2017}. We generated and downloaded light curves for each counterpart using the the web interface\footnote{\href{https://asas-sn.osu.edu}{https://asas-sn.osu.edu}}, 
using a $16\arcsec$ aperture centred on the source position, which is calibrated against the American Association of Variable Star Observers (AAVSO) Photometric All-Sky Survey \citep[APASS;][]{Henden-2012}. $V$-band photometry for both sources spans the range 2016 March 10 -- 2018 September 23 and comprises $\sim750$ measurements with an average cadence of $\sim1.25$ d. 

\section{Analysis and Results}
\label{sec:Analysis}

\subsection{IGR\,J17445$-$2747}
\label{sec:IGRJ17445_analysis}

\subsubsection{NIR Photometry}
\label{sec:IGR17445phot}
\begin{figure}
	\centering
    \includegraphics[width=0.9\columnwidth]{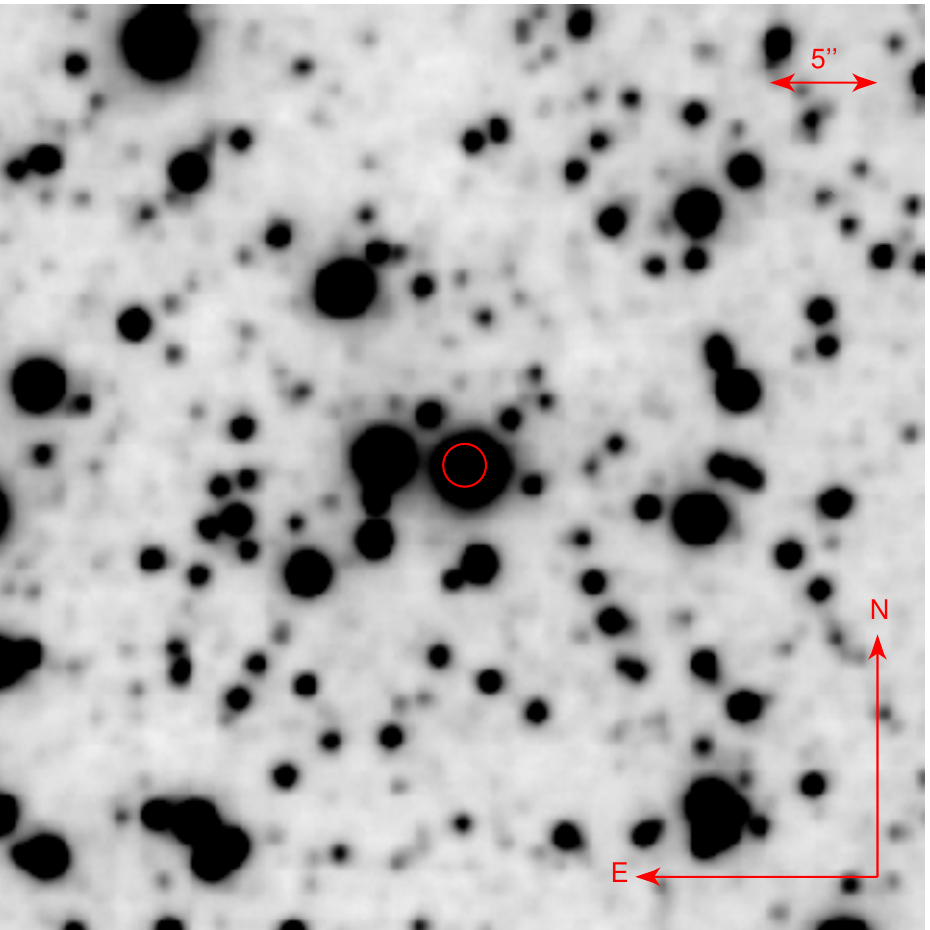}
	\caption{{\em Gemini}/NIRI $K$-band image of the IGR\,J17445$-$2747 field. The red circle represents the $1''$ radius error circle of the \cha\ X-ray position of the source \citep{Chakrabarty-2017}}
    \label{IGR17445_image}
\end{figure}

Fig.~\ref{IGR17445_image} shows the {\em Gemini}/NIRI $K$-band image of the IGR\,J17445$-$2747 field. The \cha\ X-ray position is consistent with a bright source from the Two Micron All Sky Survey \citep[2MASS;][]{Skrutskie-2006} catalogue, 2MASS\,J17443041$-$2746004, with magnitudes of $J=12.53\pm0.06$, $H=10.45\pm0.06$ and $K_s=9.65\pm0.06$.  
We measure $J=12.55\pm0.08$, $H=10.61\pm0.14$ and $K=9.69\pm0.07$ in our NIRI images. 
We found no evidence for any new outbursting source within the \cha\ error circle. The star's PSF 
is Gaussian, 
suggesting it is not  
confused.

Our NIRI magnitudes are completely consistent with those of the 2MASS catalogue. However, a lack of variability does not rule it out as the true NIR counterpart of IGR\,J17445$-$2747. It is possible that the source had decayed to NIR quiescence at the time of the NIRI observations, as they were performed five days after the detection of the X-ray outburst by {\em Swift}. However, the source was still detected in X-rays (at a 0.3--$10 {\rm \, keV}$ unabsorbed flux $F_X\sim2\times10^{-11} {\rm \, erg \, s^{-1}}$ cm$^{-2}$) at the time of the NIRI observations \citep[][in prep.]{Bahramian-2020-prep}.

We calculate the probability that 2MASS\,J17443041$-$2746004 is a chance alignment with the {\em Chandra} position to be just 0.3 per cent. We calculate this by considering the $1''$ uncertainty on the {\em Chandra} position of the X-ray source \citep{Chakrabarty-2017} and the spatial density of NIR sources as bright or brighter than 2MASS J17443041$-$2746004 (91 such sources in a $3'$ radius around the X-ray position). We note that if we consider NIR sources that are at least 1 per cent the flux of 2MASS\,J17443041$-$2746004, this probability is still as low as 3 per cent. It is therefore highly likely that this bright source is the true counterpart to IGR\,J17445$-$2747. 

\subsubsection{Optical Spectroscopy}
\label{sec:IGR17445optspec}

\begin{figure}
	\centering
    \includegraphics[width=\columnwidth]{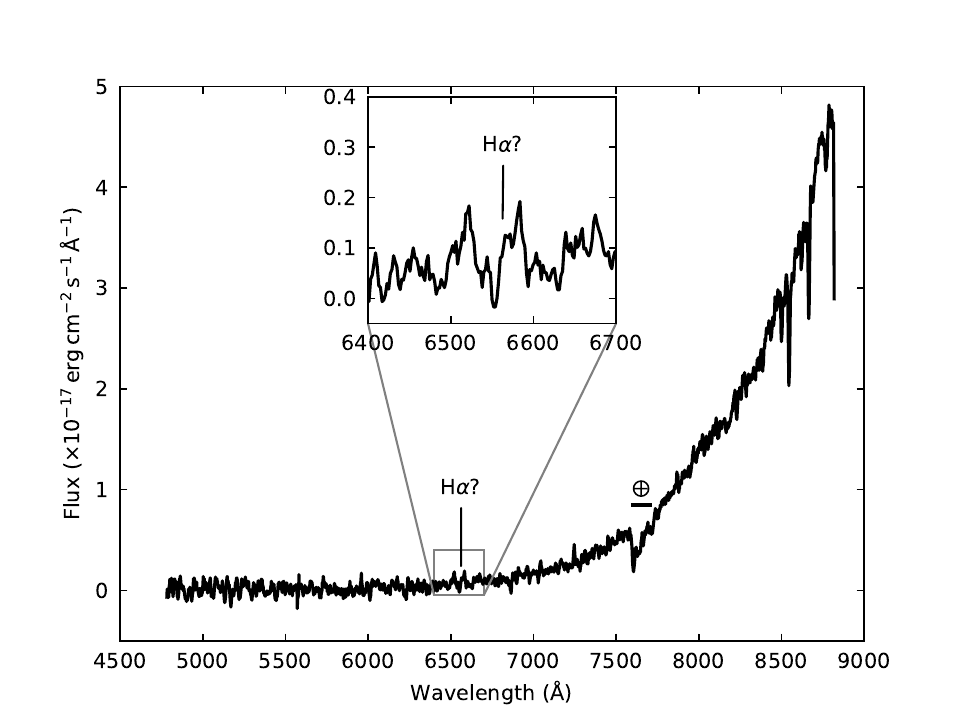}
	\caption{{\em SOAR}/Goodman spectrum of the suspected optical counterpart to IGR\,J17445$-$2747, smoothed with a 5-point boxcar function. A zoomed in portion of the region around 6562.8 \AA\ is shown inset, indicating a potential weak H$\alpha$ emission feature. The absorption feature at 7590--7720 \AA\ is due to atmospheric absorption in the ${\rm O}_2$ A-band and is labelled with the $\oplus$ symbol.}
    \label{IGR17445_opt_spec}
\end{figure}

The {\em SOAR}/Goodman spectrum of the suspected optical counterpart to IGR\,J17445$-$2747 is shown in Fig.~\ref{IGR17445_opt_spec}. The spectrum suffers from heavy extinction, with the source barely detected above the noise at wavelengths $\lesssim7000$ \AA\ in the 1800 s exposure. Considering the source is extremely bright at NIR wavelengths, its faintness in the optical regime suggests that the source is likely not a close-by dwarf star.

Highlighted in Fig.~\ref{IGR17445_opt_spec} is a possible emission feature that could be consistent with H$\alpha$ and therefore an indication of an accreting system. However, as the supposed feature is located in a region of the spectrum where pixel-to-pixel noise variations dominate, we cannot confirm its identity as a true emission line. We must instead focus on the properties of the NIR spectrum to interpret the nature of the system.

\subsubsection{NIR Spectroscopy}
\label{sec:IGR17445nirspec}

\begin{figure}
    \centering
    \begin{subfigure}[b]{0.49\textwidth}
        \includegraphics[width=\textwidth]{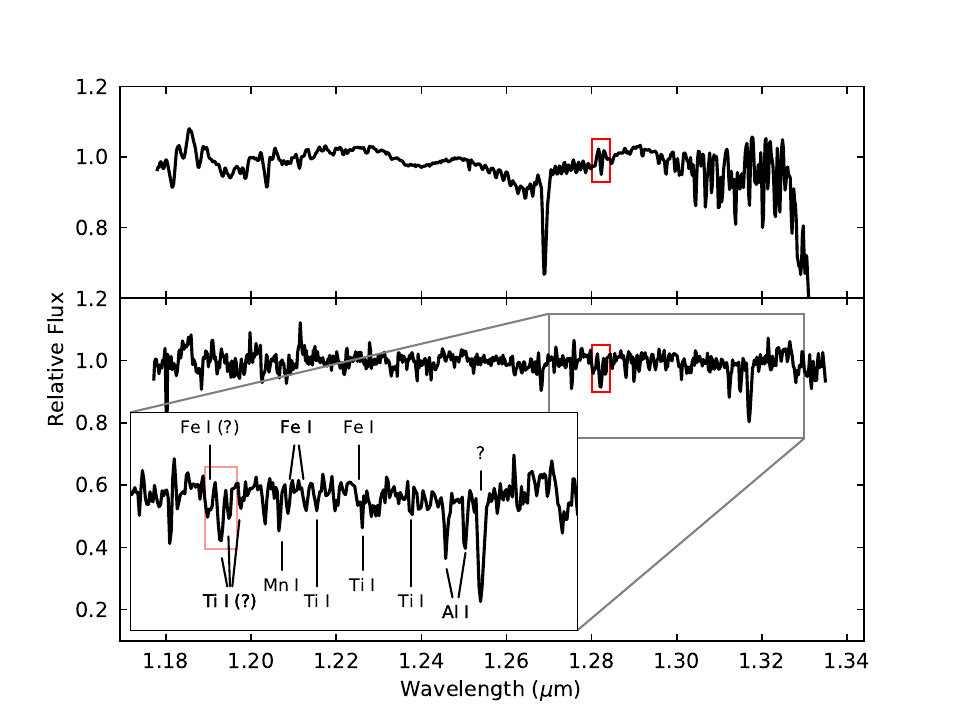}
    \vspace{-5mm}
    \end{subfigure}
    \vspace{-3mm}
    \begin{subfigure}[b]{0.49\textwidth}
        \includegraphics[width=\textwidth]{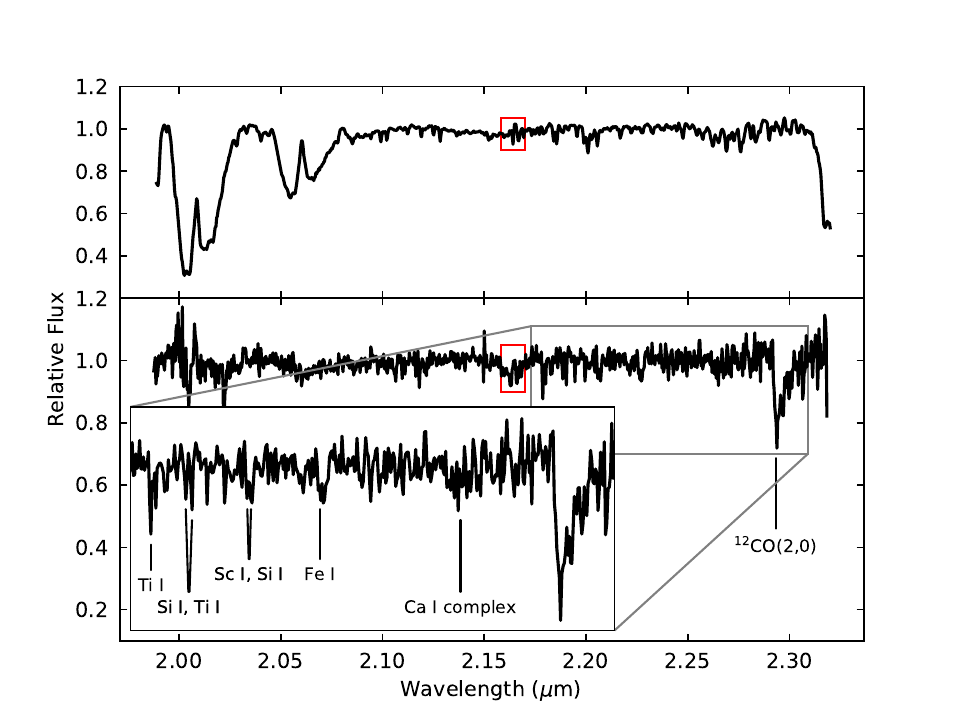}
    \end{subfigure}
    \caption{{\em Gemini}/Flamingos-2 spectra of 2MASS\,J17443041$-$2746004, the suspected NIR counterpart to IGR\,J17445$-$2747, with the $J$-band in the {\itshape top} and $K_s$-band in the {\itshape bottom} sub-figures. In both sub-figures, the upper panel shows the telluric transmission spectrum used to correct the target spectrum and the lower panel continuum normalised, telluric corrected spectrum of the target, velocity corrected to the rest frame using the heliocentric velocity $v_{\rm helio}=208 {\rm \, km \, s^{-1}}$. In both sub-figures we show a zoomed in portion of the target spectrum, with some spectral lines identified and labelled. The red boxes highlight where hydrogen lines (Pa$\beta$ in the $J$-band and Br$\gamma$ in the $K_s$-band) have been removed from the spectrum of the telluric standard, resulting in some residuals that may have affected the telluric correction. As such, any line identifications in this region are labelled as uncertain with `(?)'. The line labeled `?' in the $J$-band spectrum is an unknown absorption feature.}
    \label{IGR17445_NIR_spec}
\end{figure}

\begin{table*}
 \centering
  \caption{Definitions of the spectral features and the passbands used to estimate the continuum at each feature for two-dimensional spectral classification. We use the same definitions as \citet{Bahramian-2014b}.}
  \begin{tabular}{@{}ccccccc@{}}
  \hline
  		&		\multicolumn{2}{c}{Line}			&		\multicolumn{2}{c}{Blue continuum}	&	\multicolumn{2}{c}{Red continuum}\\
  Feature	&	Centre ($\mu$m)	&	$\delta \lambda$	&	Centre ($\mu$m)	&	$\delta \lambda$	&	Centre ($\mu$m)	&	$\delta \lambda$\\
  \hline
  Na {\sc i}	&	2.2075		&	0.007				&	2.1940		&	0.006				&	2.2150		&	0.004\\
  Ca {\sc i}	&	2.2635		&	0.011				&	2.2507		&	0.0106				&	2.2710		&	0.002\\
  $^{12}$CO &	2.2955		&	0.013				&	2.2500		&	0.016				&	2.2875		&	0.007\\
  \hline
\end{tabular}
\label{tab:specid_regions}
\end{table*}

The $J$- and $K_s$-band {\em Gemini}/Flamingos-2 spectra of 2MASS\,J17443041$-$2746004, the suspected NIR counterpart to IGR\,J17445$-$2747, are shown in Fig.~\ref{IGR17445_NIR_spec}. We identified a number of prominent lines in the spectra and used the {\tt rvidlines} {\sc iraf} task to measure the radial velocity (RV) of the source, deriving a heliocentric velocity $v_{\rm helio}=208\pm27 {\rm \, km \, s^{-1}}$ and $v_{\rm helio}=208\pm13 {\rm \, km \, s^{-1}}$ in the $J$- and $K_s$-band spectra, respectively. These velocities were applied to the relevant spectra, resulting in the Doppler corrected versions plotted in Fig.~\ref{IGR17445_NIR_spec}. We identified  spectral lines  using published spectral libraries for late-type stars \citep{Kleinmann-1986,Wallace-1997,Wallace-2000} and the National Institute of Standards and Technology - Atomic Spectra Database \citep[NIST-ASD;][]{Kramida-2018}.

The most prominent lines in the NIR spectrum are 
neutral metal species (e.g.\ Al {\sc i}, Ti {\sc i}) and the molecular $^{12}$CO (2,0) bandhead, 
typically associated with late-type (K--M) stars \citep[see e.g.][]{Kleinmann-1986,Wallace-1997,Wallace-2000}. The strong $^{12}$CO feature, combined with the brightness of the NIR counterpart and faintness of the optical counterpart provide compelling evidence that the optical/NIR counterpart is a distant late-type giant. 
We see no 
evidence for Pa$\beta$ or Br$\gamma$ emission. 

We test the giant hypothesis by adopting the approach of \citet{Bahramian-2014b} and applying two-dimensional spectral classification techniques to the $K_s$-band spectrum of IGR\,J17445$-$2747. These techniques follow the methods of \citet{Ramirez-1997,Ivanov-2004} and \citet{Comeron-2004}, who identify spectral features that are temperature dependent and compare them with features that are both temperature and surface gravity dependent. We adopt the Na {\sc i} and Ca {\sc i} passbands centred at 2.2075 and $2.2635\, \mu$m, respectively, as the temperature dependent features, and the $^{12}$CO (2,0) bandhead centred at $2.2955 \, \mu$m as the temperature and surface gravity dependent feature. For each feature we calculate the equivalent width (EW), approximating the continuum level with a best-fit 1-dimensional polynomial
({\sc astropy}'s {\tt Polynomial1D})
using two nearby, featureless regions of the spectrum. We use the same feature and continuum definitions as \citet{Bahramian-2014b}, who follow the method of \citet[][see Table \ref{tab:specid_regions}]{Comeron-2004}.

We estimated uncertainties on the reported EWs using the `bootstrap-with-replacement' technique. We determined 10000 new EWs using continuum levels 
from a random sample of the 
passbands described in Table \ref{tab:specid_regions} and computed the standard error. For each new continuum passband, we allowed the same wavelength-flux pair to be selected multiple times (hence `replacement'). The EWs and 
uncertainties are reported in Table \ref{tab:IGRJ17445_EWs}, and 
discussed 
in Section \ref{sec:IGR17445_discussion}.

\begin{table}
    \centering
    \caption{Equivalent widths of spectral features in the {\em Gemini}/Flamingos-2 $K_s$-band spectrum of the NIR counterpart to IGR\,J17445$-$2747}
    \begin{tabular}{c|c}
    \hline
    Feature & EW (\AA)  \\
    \hline
    Na {\sc i} & $\phn1.21\pm0.27$\\
    Ca {\sc i} & $\phn0.66\pm0.95$\\
    $^{12}$CO  & $13.50\pm0.88$\\
    \hline
    \end{tabular}
    \label{tab:IGRJ17445_EWs}
\end{table}

\subsection{Swift\,J175233.9$-$290952}
\label{sec:SwiftJ175233_analysis}

\subsubsection{NIR imaging}
\label{sec:SwiftJ175233phot}

\begin{figure}
    \centering
    \begin{subfigure}[b]{0.45\textwidth}
        \includegraphics[width=\textwidth]{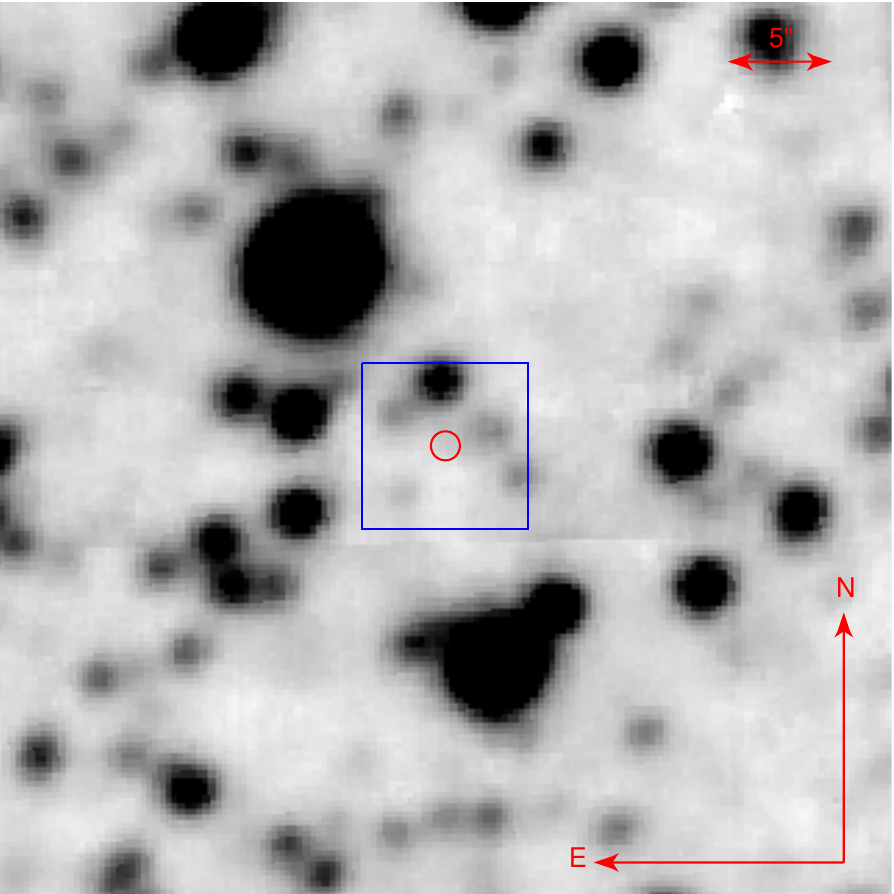}
    \end{subfigure}
    \begin{subfigure}[b]{0.45\textwidth}
        \includegraphics[width=\textwidth]{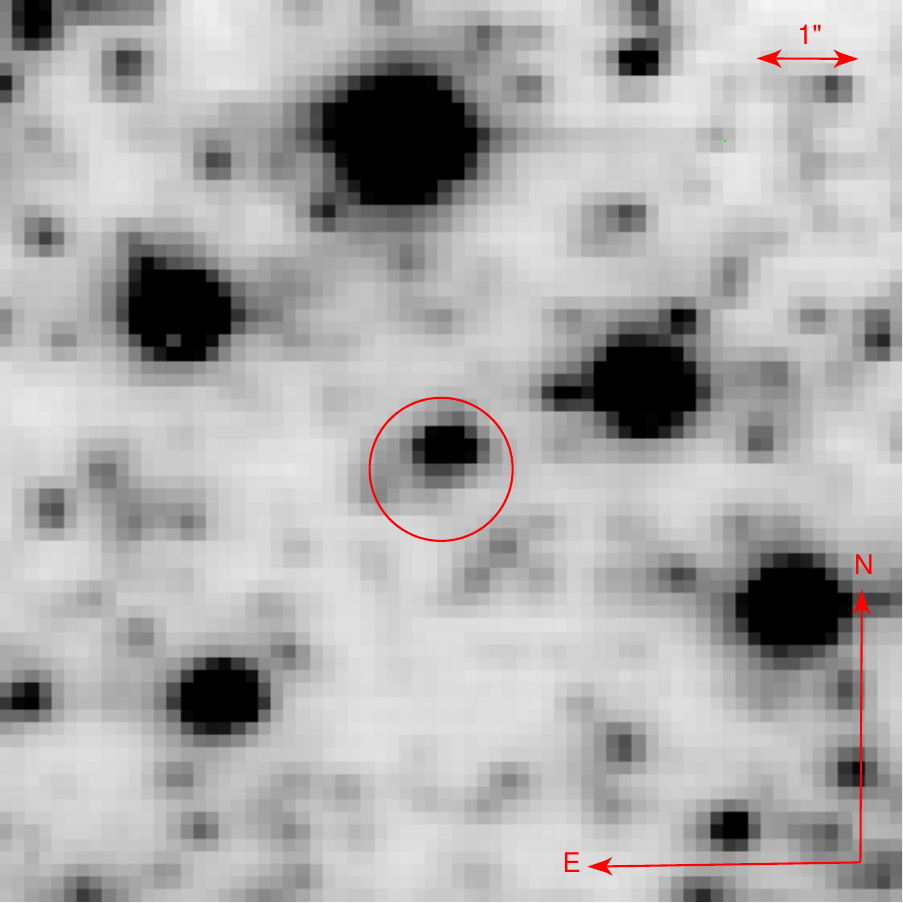}
    \end{subfigure}
    \caption{Images of the Swift\,J175233.9$-$290952 field. The {\itshape top} sub-figure shows GROND $K$-band image, observed five days after the detection of the X-ray source. The blue box represents the approximate FOV of the {\em VLT}/SINFONI spectroscopic follow-up, the median cube image (the full data cube compressed on the spectral axis) of which is shown in the {\itshape bottom} sub-figure. In both sub-figures the red circle represents the $0.7''$ radius error circle of the \cha\ X-ray position of the source.}
    \label{SwiftJ175233_images}
\end{figure}

Fig.~\ref{SwiftJ175233_images} shows the GROND $K$-band image of the Swift\,J175233.9$-$290952 field. There is no source detected inside the \cha\ error circle. The $3\sigma$ limiting magnitudes of the GROND images are $K<14.6$, $H<14.8$ and $J<15.8$. The proposed counterpart, VVV\,J175233.93$-$290947.66, has 
$J=16.4\pm0.1$ and $H=15.1\pm0.1$, so had not brightened significantly 
at the time of the GROND observations, 
five days after the initial
Swift detection. The source was still X-ray active at the time of the GROND observations \citep[][in prep.]{Bahramian-2020-prep} but it is possible that a NIR outburst could have decayed by this point.

\subsubsection{NIR spectroscopy}
\label{sec:SwiftJ175233spec}

The {\em VLT}/SINFONI median cube image 
of the Swift\,J175233.9$-$290952 field, obtained 49 d after the GROND imaging observation, is also shown in Fig.~\ref{SwiftJ175233_images} and shows the suspected counterpart, VVV\,J175233.93$-$290947.66, in the \cha\ error circle. We calculate a chance alignment probability of 8 per cent. We studied the VVV light curve of this source\footnote{generated from the  detection table at the source position, \href{http://horus.roe.ac.uk:8080/vdfs/Vregion_form.jsp}{http://horus.roe.ac.uk:8080/vdfs/Vregion\_form.jsp}} to search for outbursting behaviour or other variability, which may help us identify it as the correct counterpart. However, aside from an apparent $\sim1$ magnitude drop at MJD $\sim55309$ that 
is likely 
poor calibration (as the same pattern appears in comparison stars), there was no variability.

\begin{figure}
	\centering
    \includegraphics[width=\columnwidth]{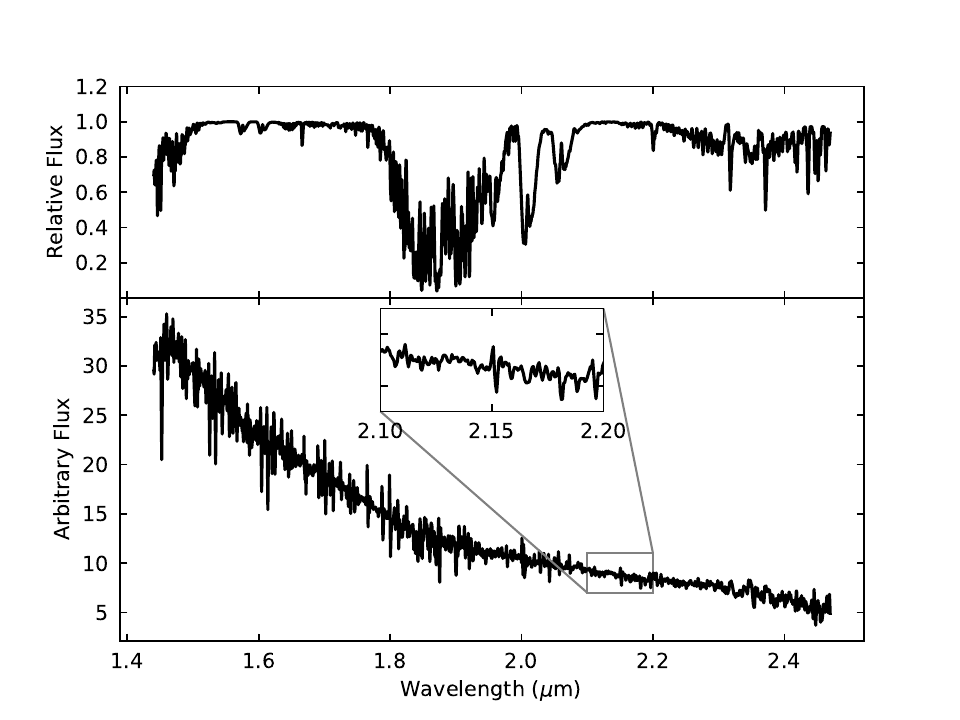}
	\caption{{\em VLT}/SINFONI $H$+$K$-band spectrum of VVV\,J175233.93$-$290947.66, the suspected NIR counterpart to Swift\,J175233.9$-$290952. The upper panel shows the telluric transmission spectrum derived with {\sc molecfit} and used to correct the target spectrum. The lower panel shows the telluric corrected spectrum of the target and has not been continuum normalised. The lower panel inset shows a zoomed in portion of the target spectrum around the rest wavelength of Br$\gamma$, highlighting the absence of emission features in this spectral region.}
    \label{SwiftJ175233_spec}
\end{figure}

The spectrum of VVV\,J175233.93$-$290947.66, shown in Fig.~\ref{SwiftJ175233_spec}, is strikingly featureless and dominated by noise (particularly in the $H$-band), with no evidence for hydrogen emission at Br$\gamma$, or any other features common in accreting binaries. There are also no absorption features suggestive of a late-type star (e.g.\ CO bandheads), which, 
with the lack of an optical counterpart, rules out a close-by flaring dwarf. 

A second, fainter source lies partially inside the \cha\ error circle in Fig.~\ref{SwiftJ175233_images}, which is not present in any NIR source catalogues, making the identification of the true counterpart less certain. Through comparison with the catalogued sources in the image, we estimate $K\sim18$ for the fainter source. 
We extract a low S/N spectrum, 
to search for 
strong accretion signatures, 
but 
do not find any evidence of Br$\gamma$. 
With the two possible counterparts in mind, we discuss the  nature of Swift\,J175233.9$-$290952 in Section \ref{sec:SwiftJ175233_discussion}.

\subsection{1SXPS\,J174215.0$-$291453}
\label{sec:SwiftJ1742analysis}

\subsubsection{Long-term Optical Monitoring}
ASAS-SN has monitored {\em Gaia} DR2
4057126472597377152 \citep{Gaia-2016,Gaia-2018b}, the optical counterpart to 1SXPS\,J174215.0$-$291453, for $\sim2.5 \, {\rm y}$. We performed timing analysis on the $V$-band light curve to search for periodicities, $P$, that may
identify a binary orbital period ($P_{\rm orb}$).
Periodicities are determined using a Lomb-Scargle periodogram analysis \citep{Lomb-1976,Scargle-1982}, which utilizes least-squares fitting of sinusoids to the light curve data to determine the power at each frequency. Lomb-Scargle analysis was performed using 
{\sc astropy} \citep{Astropy-2013,Astropy-2018}.

\begin{figure}
	\centering
    \includegraphics[width=\columnwidth]{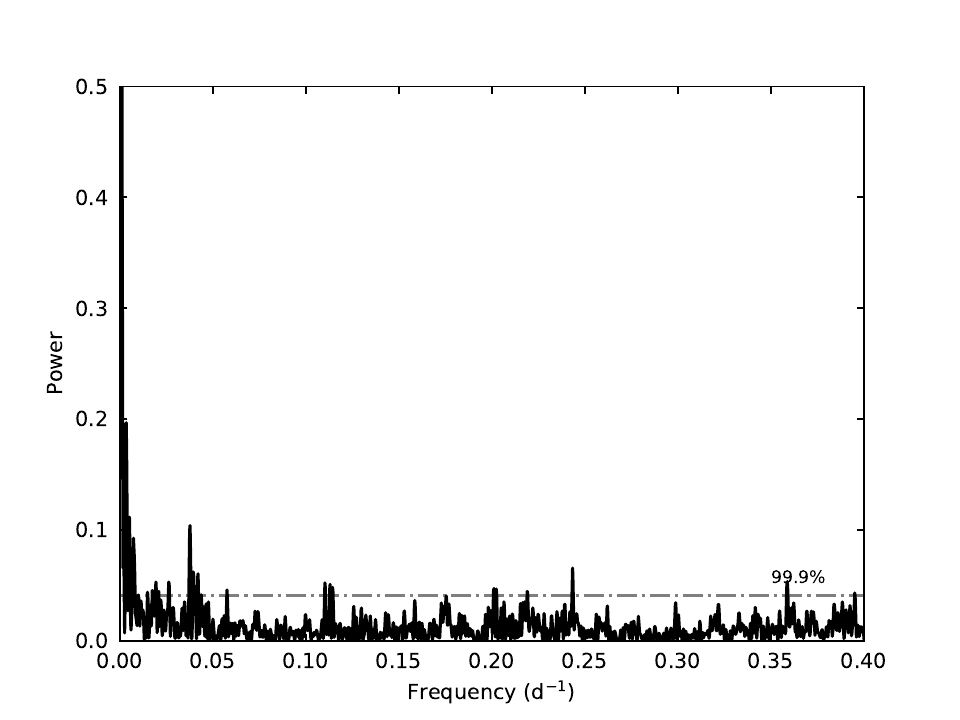}
	\caption{Lomb-Scargle periodogram, calculated using the ASAS-SN $V$-band photometry of the optical counterpart to 1SXPS\,J174215.0$-$291453. The dash-dotted line represents the 99.9 per cent significance power.}
    \label{SwiftJ1742_ASAS-SN_periodogram}
\end{figure}

The periodogram is shown in Fig.~\ref{SwiftJ1742_ASAS-SN_periodogram}. The significance of the power (represented in Fig.~\ref{SwiftJ1742_ASAS-SN_periodogram} by a dash-dotted line at 99.9 per cent confidence) is calculated by randomly shuffling the light curve magnitudes but keeping the time-stamps the same, in essence creating a randomized light curve but with the same sampling as the original data. This algorithm is excellent for identifying the peaks which are significant relative to a white noise power spectrum, but it can identify apparently significant peaks which are not significant relative to a red noise power spectrum; where statistically significant variability is detected, it is likely to be real variability, but further work must be done to determine whether the variability is periodic or aperiodic.  The peak Lomb-Scargle powers were recorded for 10000 of the random light curves and the significance percentiles derived. The highest peak suggests a periodicity of $P=926 {\, \rm d}$, which is the length of the light curve, so 
simply indicates a long-term brightening trend.
over the course of the ASAS-SN monitoring. There are other peaks above the 99.9 per cent confidence level, but folding the light curves on these periods reveals no periodic structure, and so it is likely that these peaks are indicating red noise rather than periodic variability.
We 
do not consider 1SXPS\,J174215.0$-$291453 to exhibit a periodicity.

\subsubsection{Optical Spectroscopy}

\begin{figure}
	\centering
    \includegraphics[width=\columnwidth]{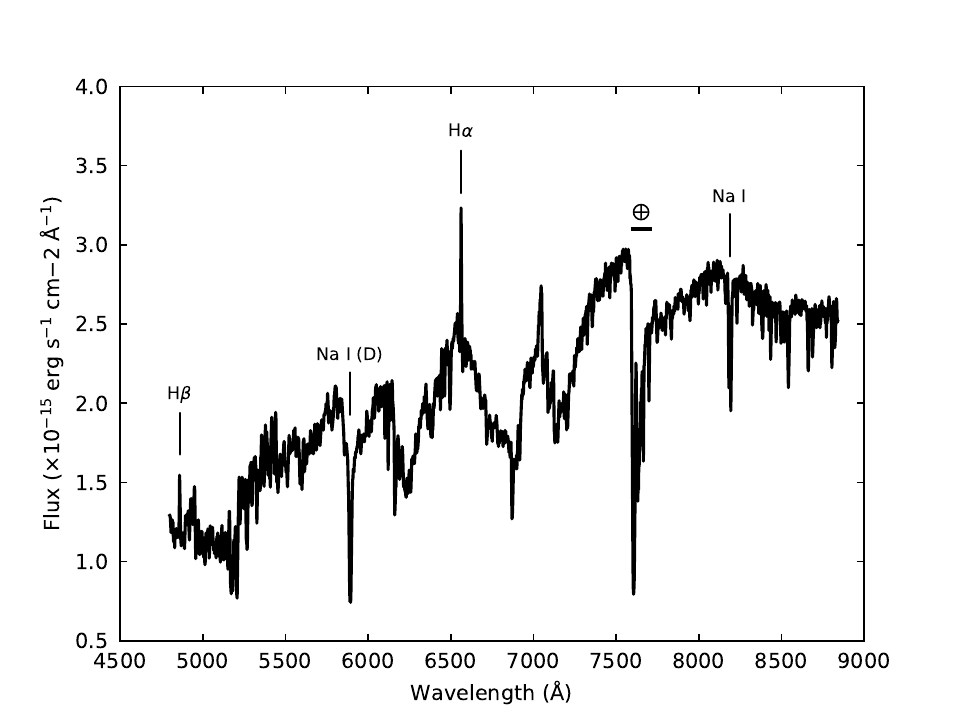}
    \vspace{-3mm}
	\caption{{\em SOAR}/Goodman spectrum of {\em Gaia} DR2 4057126472597377152, the optical counterpart to 1SXPS\,J174215.0$-$291453 (=3XMM\, J174214.9$-$29145), velocity corrected to the rest frame using the heliocentric velocity $v_{\rm helio}=-104\pm2 {\rm \, km \, s^{-1}}$. Some spectral lines have been identified and labelled. The absorption feature at 7590--7720 \AA\ is due to atmospheric absorption in the ${\rm O}_2$ A-band and is labelled with the $\oplus$ symbol.}
    \label{SwiftJ1742_SOAR_spec}
\end{figure}

The {\em SOAR}/Goodman spectrum of {\em Gaia} DR2 4057126472597377152 is shown in Fig.~\ref{SwiftJ1742_SOAR_spec}. There are a number of strong absorption lines in the optical spectrum, in particular Na {\sc i}, which is intrinsic to late-type stars and likely contributes to the interstellar Na D doublet at 5890 \AA\ and 5896 \AA. In addition, the broad absorption bandheads in the range $\sim6100-7400$ \AA\ are likely due to a combination of CaH, CaOH and TiO, common in late-type (late K--M) stars \citep{Reid-1995,Hawley-1996}.

We also identified narrow Balmer emission lines, and used these along with the {\sc iraf} task {\tt rvidlines} to measure the RV of the source, deriving a heliocentric velocity $v_{\rm helio}=-104\pm2 {\rm \, km \, s^{-1}}$. This velocity was applied to the spectrum, resulting in the Doppler corrected version plotted in Fig.~\ref{SwiftJ1742_SOAR_spec}.

We can use the optical spectrum to derive the spectral type of 1SXPS\,J174215.0$-$291453, 
similarly to 
IGR\,J17445$-$2747. \citet{Reid-1995} defined a series of flux ratios to measure the molecular features in late-type stellar spectra. The ratio is defined as $R={F_{\rm W}}/{F_{\rm cont}}$, where $F_{\rm W}$ is the integrated flux of the feature, normalised by the bandwidth, and $F_{\rm cont}$ is a pseudo-continuum flux, ie. the mean flux in a defined sideband. The strongest TiO bandhead (TiO 5; $7126-7135$ \AA) serves as the most reliable spectral type indicator between K7 and M6.5, and using the range $7042-7046$ \AA\ to calculate $F_{\rm cont}$ we find $R\approx0.67$. 
We discuss the implications in Section \ref{sec:SwiftJ1742_discussion}.

\subsection{3XMM\,J174417.2$-$293944}

\subsubsection{Long-term Optical Monitoring}
\label{sec:XMMJ1744_ASAS-SN}

\begin{figure}
    \centering
    \begin{subfigure}[b]{0.45\textwidth}
        \includegraphics[width=\textwidth]{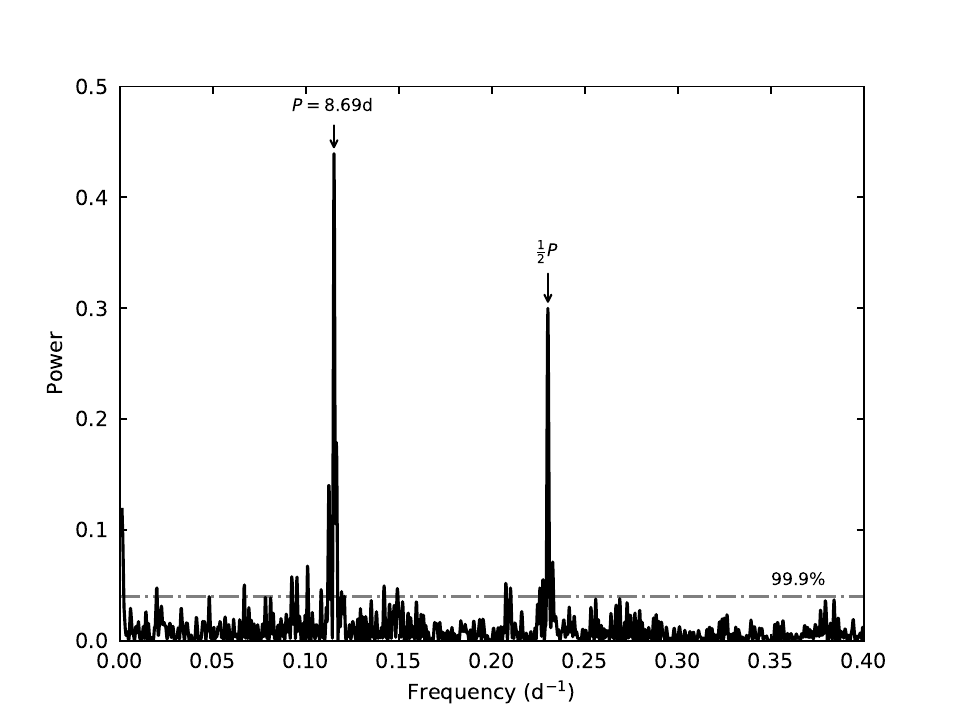}
    \vspace{-5mm}
    \end{subfigure}
    \vspace{-3mm}
    \begin{subfigure}[b]{0.45\textwidth}
        \includegraphics[width=\textwidth]{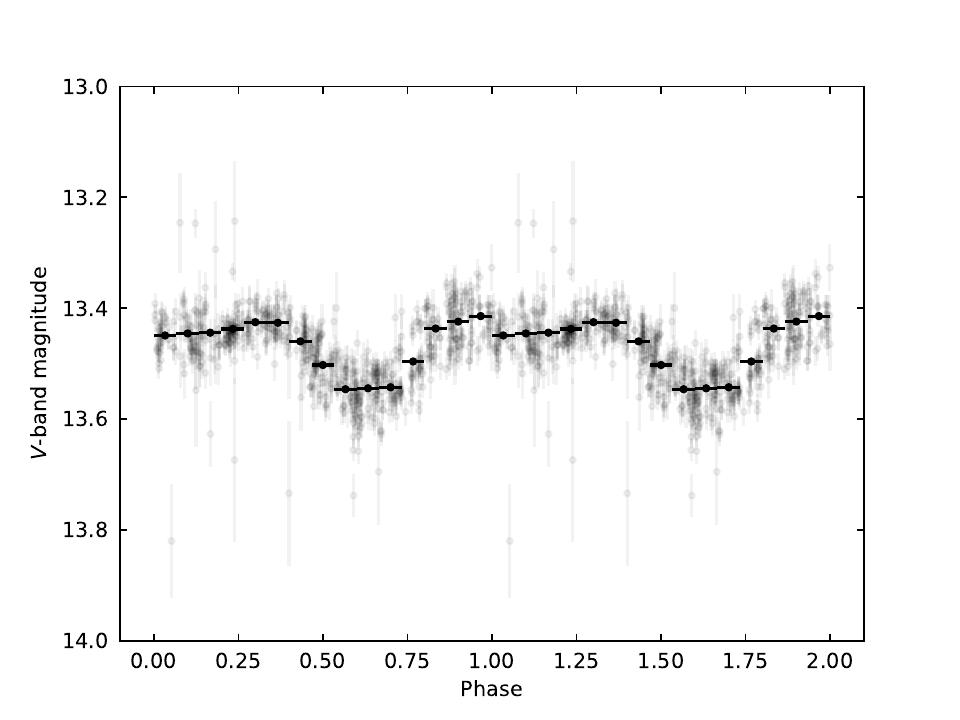}
    \end{subfigure}
    \caption{Timing analysis of the long-term $V$-band monitoring of the optical counterpart to 3XMM\,J174417.2$-$293944 with ASAS-SN. The {\itshape top} sub-figure shows the Lomb-Scargle periodogram. The periodicity, $P$, implied by the peak power is labeled along with its $\frac{1}{2}P$ alias. The dash-dotted line represents the 99.9 per cent significance power. The {\itshape bottom} sub-figure shows the ASAS-SN $V$-band light curve folded on the periodicity $P=8.69 {\, \rm d}$ inferred from the periodogram. The grey points represent the unbinned folded light curve; black points are grouped into 15 bins per cycle.}
    \label{XMMJ1744_ASAS-SN}
\end{figure}

ASAS-SN has monitored the optical counterpart to 3XMM\,J174417.2$-$293944 for $\sim2.5 {\, \rm y}$. We performed a Lomb-Scargle periodogram analysis on the $V$-band light curve to search for signatures of a potential $P_{\rm orb}$.
    
The periodogram is shown in Fig.~\ref{XMMJ1744_ASAS-SN}, highlighting the peak power at $P=8.69\pm0.05 {\rm \, d}$, consistent with the period reported by ASAS-SN\footnote{\href{https://asas-sn.osu.edu/variables/230506}{https://asas-sn.osu.edu/variables/230506}} and, as we later report in Section \ref{sec:XMMJ1744_RV}, the $P_{\rm orb}$ derived from the RV study. The uncertainty on $P$ was determined with the `bootstrap-with-replacement' technique, whereby the light curve is reconstructed using a random sample of the original ASAS-SN light curve, allowing each time-flux pair to be selected multiple times.  
10000 of these light curves are subjected to Lomb-Scargle periodogram analysis and the standard deviation of the peak periodicities are used to determine the uncertainty on $P$. The significance of the power was calculated in the same way as for 1SXPS\,J174215.0$-$291453.

The ASAS-SN $V$-band lightcurve, folded on the optimal period $P=8.69 {\, \rm d}$ implied by the Lomb Scargle analysis, is also presented in Fig.~\ref{XMMJ1744_ASAS-SN}. We choose the reference time $T_0=2458334.9450$ (BJD), derived from the RV study (see Section \ref{sec:XMMJ1744_RV}) when constructing the folded light curve so we 
can directly compare to 
 the RV curve. The folded light curve shows a distinct modulation, reminiscent of orbital variability seen in binary systems and, as we  show in Section \ref{sec:XMMJ1744_RV}, $P=8.69\pm0.05 {\rm \, d}$ is completely consistent with the orbital period derived from the RV curve. We discuss the  nature of the photometric variability in Section \ref{sec:XMMJ1744_discussion}.

\subsubsection{NIR spectroscopy}

\begin{figure}
    \centering
    \begin{subfigure}[b]{0.45\textwidth}
        \includegraphics[width=\textwidth]{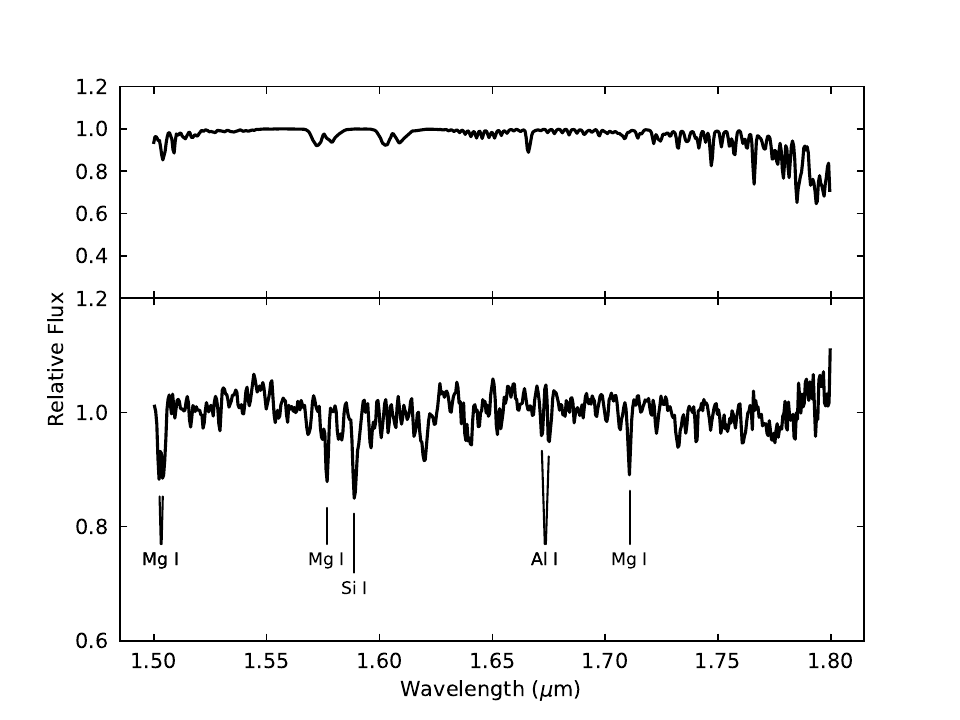}
    \vspace{-5mm}
    \end{subfigure}
    \vspace{-3mm}
    \begin{subfigure}[b]{0.45\textwidth}
        \includegraphics[width=\textwidth]{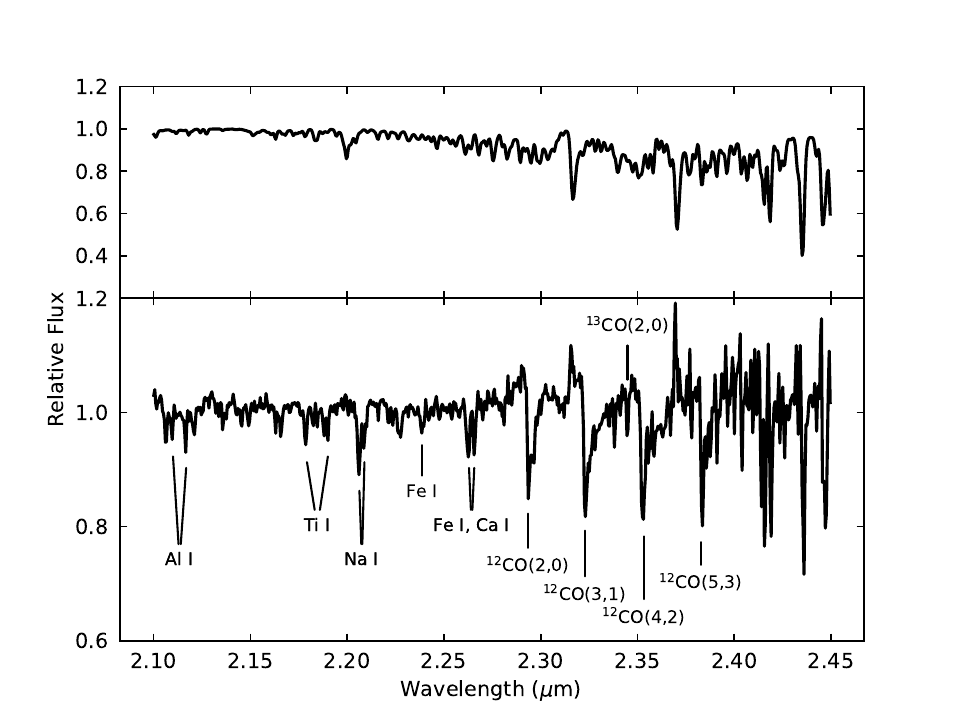}
    \end{subfigure}
    \caption{{\em VLT}/SINFONI spectra of the NIR counterpart to 3XMM\,J174417.2$-$293944 with the $H$-band portion of the spectrum in the {\itshape top} and $K$-band portion of the spectrum in the {\itshape bottom} sub-figure. In both sub-figures, the upper panel shows the telluric transmission spectrum derived with {\sc molecfit} and used to correct the target spectrum and the lower panel shows the continuum normalised, telluric corrected spectrum of the target. The $H$-band spectrum has been velocity corrected to the rest frame using the heliocentric velocity $v_{\rm helio}=134\pm7 {\rm \, km \, s^{-1}}$ and the $K$-band spectrum corrected using $v_{\rm helio}=45\pm {\rm \, km \, s^{-1}}$. Some spectral lines have  been  identified  and  labelled in the lower panel of each sub-figure.}
    \label{XMMJ1744_H+K_spec}
\end{figure}

\begin{table}
    \centering
    \caption{Equivalent widths of spectral features in the $K$-band region of the {\em VLT}/SINFONI spectrum of the NIR counterpart to 3XMM\,J174417.2$-$293944.}
    \begin{tabular}{c|c}
    \hline
    Feature & EW (\AA)  \\
    \hline
    Na {\sc i} & $3.33\pm0.14$\\
    Ca {\sc i} & $2.27\pm0.38$\\
    $^{12}$CO  & $7.85\pm0.57$\\
    \hline
    \end{tabular}
    \label{tab:XMMJ1744_EWs}
\end{table}

The {\em VLT}/SINFONI spectrum of the NIR counterpart to 3XMM\,J174417.2$-$293944 has been split into $H$- and $K$-band spectra and plotted in Fig.~\ref{XMMJ1744_H+K_spec}. Similar to IGR\,J17445$-$2747, there are a number of strong absorption features present in the NIR spectrum typical of late-type stars \citep[see e.g.][]{Kleinmann-1986,Wallace-1997,Meyer-1998}. We used these features to derive $v_{\rm helio}$ using the {\sc iraf} task {\tt rvidlines}. Using only lines in the $H$-band region of the spectrum, we derive $v_{\rm helio}=134\pm14 {\rm \, km \, s^{-1}}$, whereas using only lines in the $K$-band region of the spectrum, we derive $v_{\rm helio}=45\pm7 {\rm \, km \, s^{-1}}$. Aside from being inconsistent between the two bands, these velocities also differ from the predicted velocity from the optical RV curve presented in the following section. The discrepancy between the $H$ and $K$ bands suggests a relative wavelength calibration error $\sim6$ \AA, which is approximately twice the expected RMS uncertainty of the wavelength calibration. However, this level of error is not significant enough to prevent line identification or calculation of EWs (the latter has precision that is two orders of magnitude larger than this calibration issue). Therefore we use the derived $v_{\rm helio}$ in each band to act as a re-calibration of the wavelength solution for the purposes of displaying rest-frame spectra in Fig. \ref{XMMJ1744_H+K_spec}.

Similar to our analysis of IGR\,J17445$-$2747, we can apply two-dimensional spectral classification techniques to determine the nature of the NIR counterpart to 3XMM\,J174417.2$-$293944. We again adopt the passbands and continua defined in Table \ref{tab:specid_regions} to calculate EWs. Uncertainties on EWs were estimated using the `bootstrap-with-replacement' technique described in Section \ref{sec:IGR17445nirspec}. The calculated EWs are listed in Table \ref{tab:XMMJ1744_EWs}.

\subsubsection{Optical Spectroscopy and Radial Velocities}
\label{sec:XMMJ1744_RV}

\begin{figure}
    \centering
    \begin{subfigure}[b]{0.45\textwidth}
        \includegraphics[width=\textwidth]{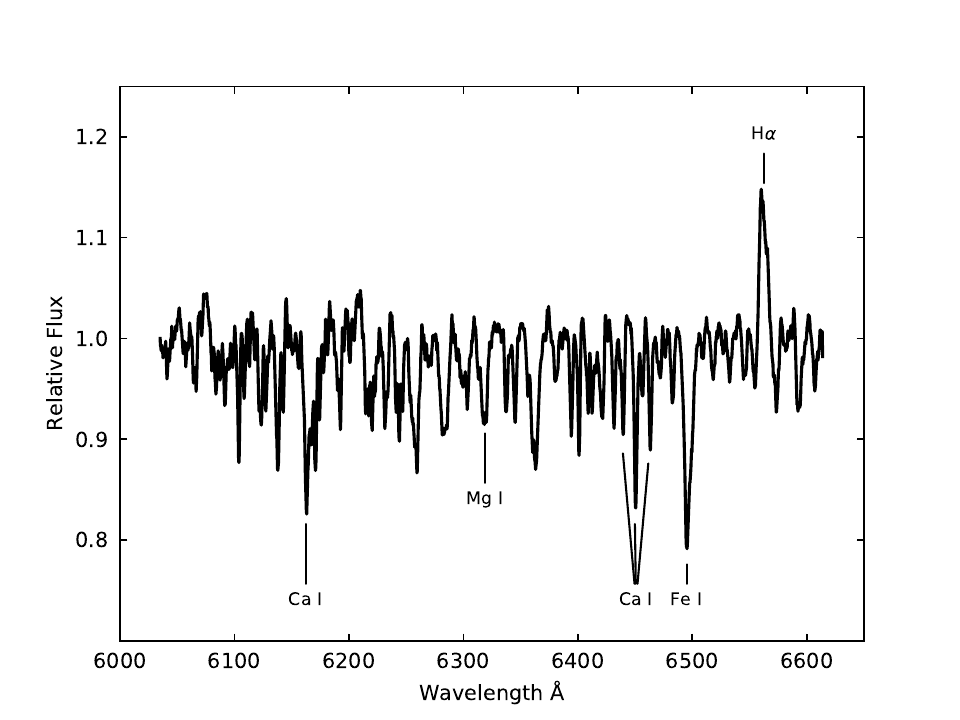}
    \vspace{-5mm}
    \end{subfigure}
    \vspace{-3mm}
    \begin{subfigure}[b]{0.45\textwidth}
        \includegraphics[width=\textwidth]{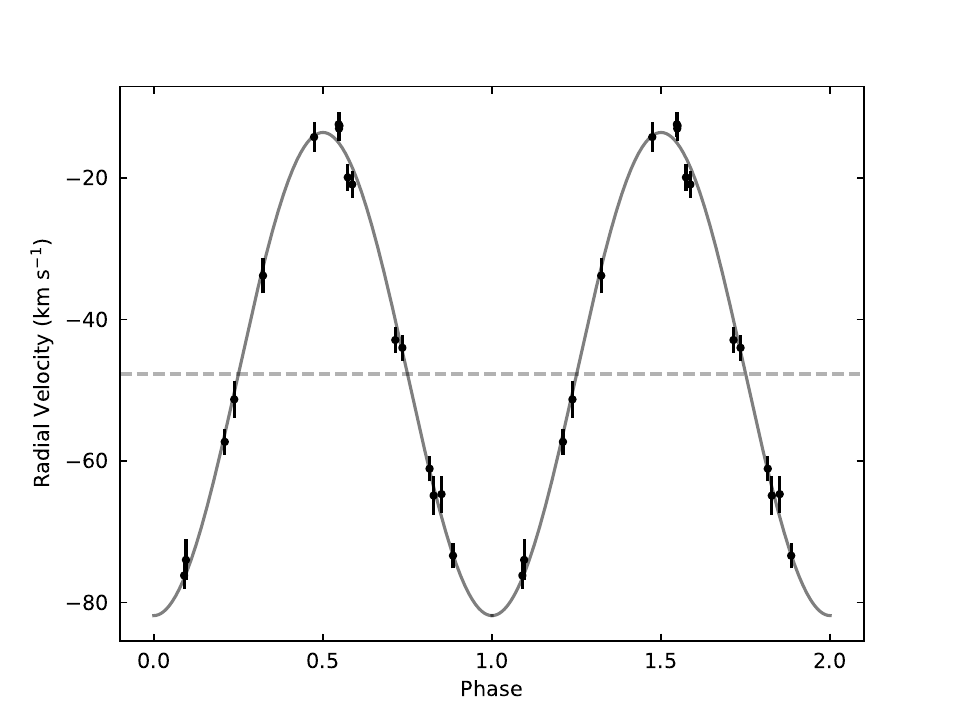}
    \end{subfigure}
    \caption{Radial velocity study of the optical counterpart to 3XMM\,J174417.2$-$293944.  Top:  average {\em SOAR}/Goodman spectrum, continuum normalised. 
     H$\alpha$ 
     is labelled, along with some example absorption features.  Bottom: RV curve showing the motion of the secondary star in  3XMM\,J174417.2$-$293944, folded on the best-fit $P_{\rm orb} = 8.7092 {\rm \, d}$. The solid curve represents the best-fit Keplerian orbital model, with the parameters derived in the text. The dashed line denotes the systemic velocity of the system $\gamma = -47.7\pm0.5 {\rm \, km \, s^{-1}}$.}
    \label{XMMJ1744_RV}
\end{figure}

The average {\em SOAR}/Goodman optical spectrum of 3XMM\,J174417.2$-$293944 is shown in Fig.~\ref{XMMJ1744_RV}. As with the NIR spectrum, strong absorption features typical of late-type stars are present. We also note a strong, broad (FWHM $\sim400 {\rm \, km \, s^{-1}}$) H$\alpha$ emission feature, indicating that 3XMM\,J174417.2$-$293944 is probably an accreting compact object. The absorption features 
most likely originate from the binary companion. We 
used
the multiple spectra to perform a RV study of the secondary star and derive the binary parameters.

To estimate RVs, we cross-correlated the spectra with the K giant spectral standard. Given the long period of the binary (see Section \ref{sec:XMMJ1744_ASAS-SN} and below), for those epochs where two spectra of 3XMM\,J174417.2$-$293944 were obtained, we use the weighted average of the two velocities. The barycentric velocities are given in Table \ref{tab:XMMJ1744_RVs}. The corresponding dates are given as Barycentric Julian Dates (BJD) on the TDB system \citep{Eastman-2010}.

\begin{table}
    \centering
    \caption{Radial velocities obtained from cross-correlation of optical spectra of 3XMM\,J174417.2$-$293944 with a K giant spectral standard.}
    \begin{tabular}{c|c}
    \hline
BJD & RV (km s$^{-1}$) \\
\hline
2458202.8039596 & -64.9$\pm$2.7 \\
2458223.7970651 & -51.3$\pm$2.6 \\
2458243.9046060 & -12.4$\pm$1.8 \\
2458243.9145226 & -13.0$\pm$1.8 \\
2458243.9244391 & -12.6$\pm$1.9 \\
2458288.9093142 & -42.9$\pm$1.8 \\
2458289.7916434 & -61.1$\pm$1.8 \\
2458309.5989186 & -76.2$\pm$1.9 \\
2458322.5218350 & -19.9$\pm$1.9 \\
2458344.4821089 & -74.0$\pm$2.9 \\
2458345.4813442 & -57.3$\pm$1.8 \\
2458346.4666366 & -33.8$\pm$2.5 \\
2458356.4906751 & -14.2$\pm$2.1 \\
2458357.4727641 & -20.9$\pm$1.9 \\
2458367.4734528 & -44.0$\pm$1.8 \\
2458368.4812511 & -64.7$\pm$2.6 \\
2458377.4867468 & -73.4$\pm$1.8 \\ 
\hline
    \end{tabular}
    \label{tab:XMMJ1744_RVs}
\end{table}

We fit a Keplerian model to these RVs using the custom Markov Chain Monte Carlo sampler \emph{TheJoker} \citep{Price-Whelan-2017}. We began with a circular fit, in which the free parameters were the orbital period $P_{\rm orb}$, epoch of the ascending node $T_0$ (given as a Barycentric Julian Date), RV semi-amplitude of the secondary $K_2$, and systemic velocity $\gamma$.  The posterior distributions are all close to Gaussian and uncorrelated, with best-fit values: $P_{\rm orb} = 8.7092\pm0.0048 {\rm \, d}$ (consistent with 
$P=8.69\pm0.05 {\rm \, d}$ derived from the 
optical photometry), $T_0 = 2458334.9450\pm0.0680 {\rm \, d}$, $K_2 = 34.2\pm0.7 {\rm \, km \, s^{-1}}$, and $\gamma = -47.7\pm0.5 {\rm \, km \, s^{-1}}$. This fit is good: for the central best-fit values above, the RMS is only $1.8  {\rm \, km \, s^{-1}}$, and the $\chi^2$ per degrees of freedom is 14.4/13. The best-fit RV model is shown folded on $P_{\rm orb}$ in Fig.~\ref{XMMJ1744_RV}.

We also experimented with fits to eccentric orbits: the posterior distribution of the eccentricity had its mode at zero, with a median of $e=0.03$, and the RMS of this fit was slightly improved at $1.6 {\rm \, km \, s^{-1}}$. The period and semi-amplitude of this fit were identical to the circular orbit fit within their uncertainties. Given this marginal evidence for eccentricity, we adopt the circular parameters for the remainder of the paper, but note that a very small eccentricity is possible, and could be constrained with additional RV measurements.

Further inspection of the data suggests a broadening of the absorption line profiles. This is common in close binary systems and is typically due to a spin-up of the star due to tidal synchronization. We used the higher resolution spectra obtained on 2018 Aug 13 (with $R \sim 12000$) to measure $v_{r} \, \textrm{sin} \, i$, where $v_r$ is the rotational velocity of the companion and $i$ is the binary inclination. This was done as described by \citet{Strader-2014}: we obtained spectra of non-rotating stars of a similar spectral type with the same setup, and then convolved these with kernels reflecting a range of $v_{r} \, \textrm{sin} \, i$ values with a standard limb-darkening law. Cross-correlation of these convolved spectra with un-broadened spectra then gives a relation between the FWHM of the cross-correlation peak and $v_{r} \, \textrm{sin} \, i$. Using this method, we find $v_{r} \, \textrm{sin} \, i = 53.0\pm2.0 {\rm \, km \, s^{-1}}$.

\section{Discussion}
\label{sec:Discussion}

\subsection{IGR\,J17445$-$2747: A bursting neutron star with a giant companion}
\label{sec:IGR17445_discussion}
The {\em Chandra} position of IGR\,J17445$-$2747 contains one extremely bright star, which we have shown has a 0.3 per cent probability to occur by chance alignment, so we take this as the likely stellar counterpart. The faintness of this object in the optical, coupled with its brightness at NIR wavelengths, suggests that the companion is likely a distant, reddened giant star. The statistical uncertainties on the distance from the {\em Gaia} DR2 data  \citep[allowing for $1.1<d<7.6$ kpc;][]{Bailer-Jones-2018} are too large to allow a luminosity estimate sufficiently accurate to distinguish between a dwarf or giant nature, especially given that there is likely to be some substantial accretion light component in the optical bandpass. Table \ref{tab:IGRJ17445_EWs} details the EWs of NIR spectral features important for spectral classification. \citet{Comeron-2004} show (in their figs. 8-13) that the EW of the $^{12}$CO feature is always $>25$ \AA\ in supergiants. We measure EW[$^{12}$CO] = $13.50\pm0.88$\AA, more typical of giants and dwarfs. We can therefore confidently rule out the possibility of a supergiant.

It has been shown that we can separate giants from dwarfs by comparing the EWs of Na {\sc i}, Ca {\sc i} and $^{12}$CO. \citet{Ramirez-1997} demonstrate that $\log$\{EW[$^{12}$CO]/(EW[Na {\sc i}]$+$EW[Ca {\sc i}])\} should be in the range $-0.22$--0.06 for dwarfs and 0.37--0.61 for giants (see their fig.~11). We measure this quantity to be $0.86\pm0.23$, significantly above the dwarfs and likely indicative of a giant (though not completely consistent within $1\sigma$ uncertainties). A bright giant, class II, is also a possibility. A final piece of evidence to support the giant hypothesis lies in fig.~10 of \citet{Ramirez-1997}, which demonstrates that giants and dwarfs are clearly distinguished in plots of EW[Na {\sc i}] vs EW[$^{12}$CO] and EW[Na {\sc i} + Ca {\sc i}] vs EW[$^{12}$CO]. In both cases, the likely counterpart for IGR\,J17445$-$2747 lies in the portion of the plots occupied by giants. \citet{Ramirez-1997} also define a relationship between EW[$^{12}$CO] and the effective temperature, $T_{\rm eff}$ for giant stars:

\begin{equation}
    T_{\rm eff} = (5019\pm79) - (68\pm4)\times {\rm EW}[^{12}{\rm CO}].
\end{equation}

We calculate $T_{\rm eff}=4100\pm110$K, propagating the uncertainty in EW[$^{12}$CO]. The relationship between $T_{\rm eff}$ and spectral type derived for giants by \citet{vanBelle-1999} indicates that $T_{\rm eff}=4100 {\rm \, K}$ is consistent with spectral type K4.3, whilst the calibration presented by \citet{Richichi-1999} suggests a consistent K4 giant. The calibration presented by \citet{Ramirez-1997} suggests K3.6. We therefore adopt a median spectral type of K4III \citep[with a reasonable range of K3-5III;][]{Richichi-1999} for the companion star of IGR\,J17445$-$2747. This implies a typical radius of $R_{\star}=45\pm9 \, R_{\odot}$ (\citealt{vanBelle-1999};see also e.g.\ \citealt{Alonso-2000}) and therefore an expected luminosity range $L\approx500\pm200 \, {\rm L}_{\odot}$.

We can also use the derived spectral type to estimate the distance to the source. K4III stars have a characteristic $V$-band absolute magnitude $M_V=0.0$ \citep{Schmidt-Kaler-1982}. This is equivalent to $M_K=-3.3$ considering the known intrinsic $(V-K)_{\rm int}$ colour for a K4III star \citep[e.g.][]{Cox-2000,Kucinskas-2005}. The observed and known intrinsic NIR colours allow us to place a reasonable constraint on extinction in the $K$-band, $A_K\sim1.2$ \citep[adopting the known conversions between extinction in the $V$-band, $A_V$ and $A_K$;][]{Cardelli-1989}. This implies $d\sim2.3$ kpc (with a scatter of $\pm0.6$ kpc considering the adopted range of K3-5III), much closer than the minimum $d\gtrsim5$ kpc derived by \citet{Mereminskiy-2017} from a spectral analysis of the X-ray burst, assuming solar abundances), though consistent with the derived {\em Gaia} DR2 range of $1.1<d<7.6$ kpc \citep{Bailer-Jones-2018}. However, as we note above, the measured $\log$\{EW[$^{12}$CO]/(EW[Na {\sc i}]$+$EW[Ca {\sc i}])\} is significantly above the accepted range for dwarfs, and slightly higher than the giants \citep{Ramirez-1997} raising the possibility that the companion of IGR\,J17445$-$2747 is instead a class II bright giant. If this is indeed the case, then from the known absolute magnitude of a K-type class II star \citep{Schmidt-Kaler-1982} we would place IGR\,J17445$-$2747 at $d\gtrsim6$ kpc, in agreement with \cite{Mereminskiy-2017}.

The strong evidence for a giant companion suggests that IGR\,J17445$-$2747 is a symbiotic X-ray binary -- a compact object accreting from the wind of a giant donor. 
Although giants can also experience Roche lobe overflow (e.g.\ the donor to the BH low-mass X-ray binary GRS 1915+105), the expected mass transfer rates will then be quite large, $\sim10^{-8} \,   {\rm \, y^{-1}}$ \citep{Vilhu-2002}, and the accretion disc should be rather large, producing long, bright outbursts \citep{Deegan-2009}, which do not resemble the observed outbursts from IGR\,J17445$-$2747. 

If it is a symbiotic X-ray binary, then IGR\,J17445$-$2747 would be the only known example of a symbiotic X-ray binary that exhibits X-ray bursts \citep[discovered by ][]{Mereminskiy-2017}. 
X-ray bursts are thought to only be produced on NSs with $B<10^{11}$ G \citep{Bildsten-1998a}, while NSs in symbiotic X-ray binaries generally have $B\sim10^{12}$ G, explained by their accreting lifetime being too small to bury their magnetic fields from their high initial fields, as often invoked to explain low fields in LMXBs \citep{Cumming-2001}. However, not all NSs are born with high fields; several young NSs in supernova remnants have $B=10^{10-11}$ G \citep{Gotthelf-2013}, and Cir X-1, a young HMXB in a supernova remnant, also shows bursts \citep{Heinz-2013}. Thus, we argue that IGR\,J17445$-$2747 is likely a symbiotic X-ray binary where the NS was born with an initially low $B$ field. A RV study would allow us to confirm the symbiotic nature, as the known symbiotic X-ray binaries exhibit $P_{\rm orb}$ between hundreds of days and several years \citep{Yungelson-2019}.

\subsection{Swift\,J175233.9$-$290952: An unidentified VFXT}
\label{sec:SwiftJ175233_discussion}

Of all the sources followed up at optical and NIR wavelengths in this work, Swift\,J175233.9$-$290952 has the most uncertain identification. Photometric studies of the suspected counterpart, VVV\,J175233.93$-$290947.66, revealed that (a) it had not brightened significantly ($\Delta H\leq$0.3, $\Delta J\leq$0.6; Fig.~\ref{SwiftJ175233_images}) five days after the X-ray source was discovered by {\em Swift}, and (b) there was no apparent variability in the VVV source. This has rendered us unable to confirm VVV\,J175233.93$-$290947.66 as the true counterpart to Swift\,J175233.9$-$290952.

Nevertheless, we obtained NIR spectroscopy of the VVV source as it is the only suspected counterpart bright enough to study, and we are able to present some potential scenarios for the nature of Swift\,J175233.9$-$290952. We can also rule out a flaring dwarf nature for the X-ray source as it was X-ray active for 2--3 weeks after the initial detection \citep[][in prep.]{Maccarone-2017b,Bahramian-2020-prep}. X-ray flares from late-type stars are typically $\sim$minutes to $\sim$hours in duration \citep[see e.g.][]{Pye-2015}.

If the VVV source is the true counterpart to the X-ray source, there are limited possibilities for its nature. 
The lack of an optical counterpart to Swift\,J175233.9$-$290952 suggests that the source is either intrinsically red, or heavily reddened. Examining the {\em Swift}/XRT X-ray spectrum of the source we measure a neutral hydrogen column density $N_{\rm H}<7.2\times10^{21}$ cm$^{-2}$, which, utilizing the known correlation $N_{\rm H}=(2.81\pm0.13)\times10^{21}A_V$ \citep{Bahramian-2015}, equates to $A_V<2.56$. The NIR source has an observed NIR colour $J-H=1.32\pm0.13$, which, if we adopt the maximum value of $A_V$ derived from the X-ray spectrum, would imply an intrinsic $(J-H)_{\rm int}=1.05\pm0.13$ \citep[adopting the known conversions between $A_V$, $A_J$ and $A_H$;][]{Cardelli-1989}. This implies that VVV\,J175233.93$-$290947.66 is intrinsically quite red.

Could 
VVV\,J175233.93$-$290947.66  be the companion in an X-ray binary? The lack of hydrogen emission lines would be quite unusual for an X-ray binary with a hydrogen-rich accretion disc in quiescence (though some show featureless spectra in outburst, e.g.\ \citealt{Shahbaz-1996}). One may consider an ultracompact X-ray binary, which would not have H lines. The absolute magnitude of  ultracompact X-ray binaries in outburst can be estimated by observations of ultracompact X-ray binaries in globular clusters to be around $M_B\sim5\pm1$ \citep{Deutsch-2000,Haurberg-2010,Edmonds-2003a}, consistent with predictions from \citet{vanParadijs-1994}. Transient systems have rarely been detected in quiescence, due to their faintness; \citet{DAvanzo-2009} find one ultracompact with $M_V\sim13$. For a distance of 8 kpc and $A_V\sim2$ (see above), these suggest $V\sim21.5$, $V\sim30$, or (assuming $(V-K)\sim0$) $K\sim19.5$, $K\sim27.5$ for outburst and quiescence respectively. This is far too faint for VVV\,J175233.93$-$290947.66 ($H=15.1$).

Swift\,J175233.9-290952 could be a young stellar object (YSO), as these exhibit X-ray emission \citep{Feigelson-1999,Preibisch-2005}  are intrinsically red, and are often obscured by the molecular clouds in which they are born. However, NIR spectra of YSOs typically exhibit Br$\gamma$ and H$_2$ emission in the $K$-band \citep{Cooper-2013}, which is absent from VVV\,J175233.93$-$290947.66. Additionally, YSOs typically exhibit excess mid-IR emission \citep[e.g.][]{Koenig-2014}, but 
Swift\,J175233.9$-$290952 is not present in mid-IR survey catalogues such as GLIMPSE.

Swift\,J175233.9$-$290952 could be a background Active Galactic Nucleus (AGN). Some blazars (AGN with jets pointed toward us) show weak or no emission lines in their spectra \citep[e.g.][]{Landt-2004}, 
similar to 
Fig.~\ref{SwiftJ175233_spec}. However, blazars are strong radio sources, and the 10 GHz non-detection 
of Swift\,J175233.9$-$290952, combined with a radio/X-ray luminosity ratio much lower than typical accreting BHs \citep{TetarenkoA-2017}, provides strong evidence against this hypothesis.

Alternatively, 
VVV\,J175233.93$-$290947.66 may not be associated with Swift\,J175233.9$-$290952, considering the stellar crowding in this field. We note that the probability of the \cha\ position being coincident with a star as bright or brighter than VVV\,J175233.93$-$290947.66 is 8 per cent.
The first photometric measurements of the field (which found $K>14.6$) occurred five days after the initial X-ray outburst. The {\em VLT}/SINFONI observation, which reached a depth of $K\sim18$, took place more than a month later, at which point the X-ray emission had faded into quiescence \citep[][in prep.]{Bahramian-2020-prep}. As discussed above, an ultracompact X-ray binary outburst (e.g.\ $L_X\sim10^{36} {\rm \, erg \, s^{-1}}$) would likely reach $K\sim19-20$ (i.e.\ undetectable in our photometric data), while an outburst from an LMXB with a $\sim$2 h orbit (such as SAX J1808.4-3658, \citealt{Roche-1998}, or NGC 6652A, \citealt{Engel-2012}) would likely reach $K\sim16-18$, beyond the limiting magnitude of the GROND images, which were obtained during the outburst of Swift\,J175233.9$-$290952. In quiescence, an ultracompact X-ray binary would be quite undetectable, while an LMXB with $P_{\rm orb}\sim2$ h like SAX J1808.4-3658 ($M_I=7.3$, \citealt{Deloye-2008}) would likely be at $K\sim21-22$. 
Indeed, Fig.~\ref{SwiftJ175233_images} shows that there is another, fainter source  inside the \cha\ error circle at $K\sim18$. This could be the true NIR counterpart of Swift\,J175233.9$-$290952, but it is too faint to obtain a useful spectrum  with current technology and reasonable exposure times. Should the true counterpart of Swift\,J175233.9$-$290952 have a relatively short orbital period, such as the examples discussed above, the quiescent counterpart is likely undetected.

\subsection{1SXPS\,J174215.0$-$291453: A chromospherically active binary}
\label{sec:SwiftJ1742_discussion}

\begin{figure*}
    \centering
    \includegraphics[width=\textwidth]{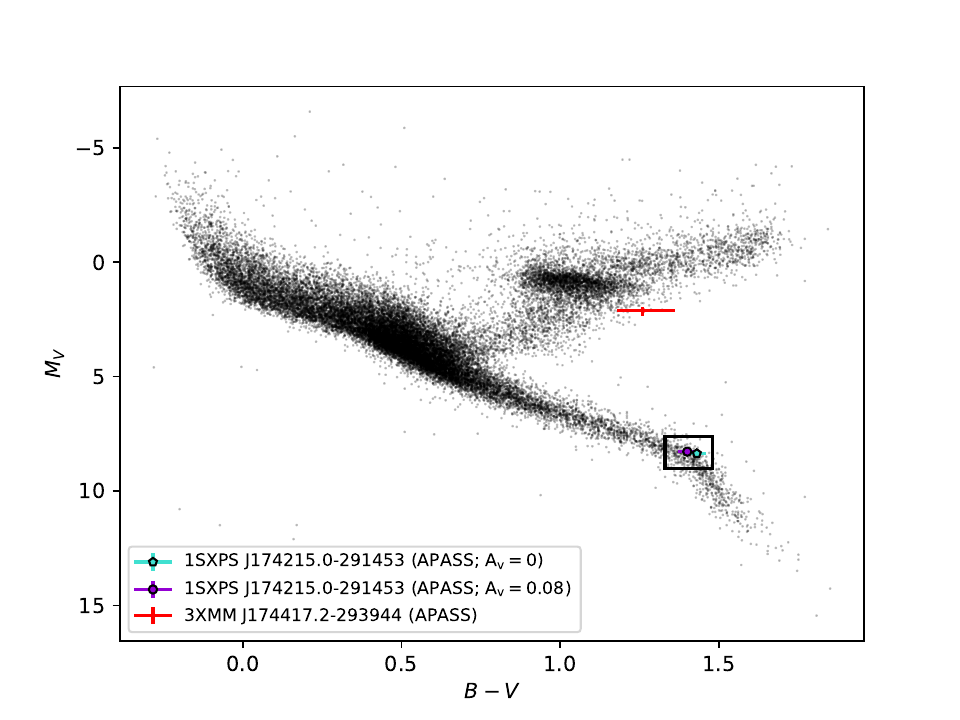}
    \vspace{-7mm}
    \caption{Hertzsprung-Russell diagram constructed from the improved {\em Hipparcos} catalogue \citep{vanLeeuwen-2007}. The y-axis denotes the absolute magnitude $M_V$ calculated from the {\em Hipparcos} parallaxes and the x-axis denotes the $B-V$ colour. The cyan and magenta points (a box is drawn to guide the eye) highlight the location of the optical counterpart to 1SXPS\,J174215.0$-$291453 on the main sequence, assuming the minimum ($A_V=0$) and maximum ($A_V=0.08$) values for extinction. 
    The red point highlights 
    the optical counterpart to 3XMM\,J174417.2$-$293944, corrected for extinction, 
    below the giant branch.
    } 
    \label{HR_diagram}
\end{figure*}

The optical counterpart of 1SXPS\,J174215.0$-$291453 has a known  {\em Gaia} distance ($d=78.1\pm0.3 {\rm \, pc}$)\footnote{from inverting the measured parallax, which is an acceptable method of deriving $d$ in cases such as this where the fractional parallax uncertainty is small \citep{Luri-2018}}, allowing us to place it on the H-R diagram.
We use the upper limit to $N_{\rm H}$ from the source's X-ray spectrum to constrain the extinction 
to $A_V<0.08$. 
Using APASS magnitudes and this extinction, we infer $M_V=8.28\pm0.02$ and an intrinsic $(B-V)_{\rm int}=1.40\pm0.03$, placing 
1SXPS\,J174215.0$-$291453 on the main sequence in a {\em Hipparcos} H-R diagram 
\citep[][see Fig.~\ref{HR_diagram}]{Perryman-1997,vanLeeuwen-2007}, classifying the source as a dwarf.

The {\em SOAR}/Goodman optical spectrum of the counterpart 
in Fig.~\ref{SwiftJ1742_SOAR_spec} exhibits lines typical of late-type stars and allows us to derive the full spectral type. 
The strong absorption bands in the range $\sim$6100--7400 \AA\ are 
due to a combination of CaH, CaOH and TiO, 
which indicate   
an M dwarf 
\citep[e.g.][]{Reid-1995}.  
Comparing the $J-K_s$ and $R-K_s$ colours \citep{Fitzpatrick-1999}, 
from 2MASS and the USNO-B1 catalogue \citep{Monet-2003}, indicates
an $\lesssim$M3V star. In \S  \ref{sec:SwiftJ1742analysis} we derive a flux ratio $R\approx0.7$ from the TiO 5 bandhead. Utilising the relationship between $R_{\rm TiO 5}$ and spectral type defined by \citet{Reid-1995}
we find that 1SXPS\,J174215.0$-$291453 is 
of 
spectral type M1$(\pm0.5)$V. This classification is supported by other flux ratios such as TiO 2 (7058-7061 \AA) and CaH 3 (6960-6990 \AA). 

The presence of narrow Balmer lines in emission indicate that the source is chromospherically active \citep{Reid-1995,Hawley-1996,Alonso-Floriano-2015}, 
explaining 
the X-ray emission detected in the {\em Swift} GBS, and by {\em XMM-Newton}. We measure 
$L_X(0.2-10\ {\rm \, keV})\sim2\times10^{29} {\rm \, erg \, s^{-1}}$ 
which 
is within the known luminosity range for M-dwarfs \citep[e.g.][]{Doyle-1989, Gonzalez-Alvarez-2019}. However, the large RV ($v_{\rm helio}=-104\pm2 {\rm \, km \, s^{-1}}$) is somewhat puzzling. {\em Gaia} DR2 measures a proper motion $\mu\approx46 {\rm \, mas \, y^{-1}}$, which translates to a transverse velocity of $\sim17  {\rm \, km \, s^{-1}}$, a normal kinematic value for a disc star. 
The large RV (compared to the transverse velocity) suggests that 1SXPS\,J174215.0$-$291453 is likely a close binary system, with at least one of the components a chromospherically active M-dwarf, and no evidence for accretion.

\subsection{3XMM\,J174417.2$-$293944: A white dwarf binary with a subgiant companion}
\label{sec:XMMJ1744_discussion}

The folded ASAS-SN $V$-band light curve of 3XMM\,J174417.2$-$293944 in Fig.~\ref{XMMJ1744_ASAS-SN} shows distinct periodic variability. 
However, the nature of the variability remains unclear. 
The light curve in Fig.~\ref{XMMJ1744_ASAS-SN} does not resemble an ellipsoidal light curve, 
which 
would have two maxima at phases 0 and 0.5, and two minima at phases 0.25 and 0.75 
(as 
we have defined orbital phase).
The variability may be due to starspots \citep[][]{Olah-2018} rotating with the star, synchronised to $P_{\rm orb}$ \citep[e.g.][]{Thompson-2018}.

The {\em VLT}/SINFONI $H+K$-band spectrum of the NIR counterpart to 3XMM\,J174417.2$-$293944 presented in Fig.~\ref{XMMJ1744_H+K_spec} 
enables us 
to identify the spectral class of the donor, similarly to IGR\,J17445$-$2747. We measure EW[$^{12}$CO] = $7.85\pm0.57$, ruling out a supergiant \citep{Comeron-2004}. We measure $\log$\{EW[$^{12}$CO]/(EW[Na {\sc i}]$+$EW[Ca {\sc i}])\}$=0.15\pm0.05$, 
between the typical values for giants and dwarfs derived by \citet{Ramirez-1997}. Comparing our measured EWs with figs. 9 and 10 and table 5 of \citet{Ramirez-1997} does little to resolve the ambiguity, with EW[$^{12}$CO] consistent with those of late-type dwarfs, but the EW[Na {\sc i}] and EW[Ca {\sc i}] are more typical of giants (though a dwarf nature is still possible with these EWs). We therefore suggest that the spectral class of 3XMM\,J174417.2$-$293944 lies between a dwarf and a giant, i.e.\ it may be a subgiant (class IV).

We derive the extinction in the direction of 3XMM\,J174417.2$-$293944, $A_V=1.37^{+0.21}_{-0.13}$ by measuring $N_{\rm H}$ from the best-fit X-ray spectrum and using the \citet{Bahramian-2015} $A_V$-$N_{\rm H}$ correlation. We calculate $M_V=2.11^{+0.23}_{-0.16}$ and $(B-V)_{\rm int}=1.26^{+0.10}_{-0.08}$ using the reported APASS magnitudes and colours and the Gaia distance $d=990\pm40 {\rm \, pc}$\footnote{from inverting the parallax, again see \citet{Luri-2018}}. These values place 3XMM\,J174417.2$-$293944 below the giant branch on the H-R diagram (Fig.~\ref{HR_diagram}). 

Assuming that the secondary (the visible star) fills its Roche lobe and that its rotation is synchronized with the orbital period, its projected rotational velocity along our line of sight is given by: 

\begin{equation}
v_{r} \, \textrm{sin} \, i = K_2 (1-e^2)^{1/2} \frac{0.49 q^{2/3} (1+q) }{0.6 q^{2/3} + \textrm{ln}(1 + q^{1/3})}.
\end{equation}

\noindent Here $q = M_2/M_1$ is the mass ratio of the secondary to the primary, and we have used the approximation for the effective radius of the donor Roche lobe from \citet{Eggleton-1983}, valid 
for all 
reasonable $q$ values. Using the posterior samples of $K_2$ and drawing samples from the observed $v_{r} \, \textrm{sin} \, i$, we find $q = M_2/M_1 = 2.40\pm0.11$. When combined with the mass function of the primary inferred from the $K_2$ and orbital period, this implies  $M_{1} \, \textrm{sin}^3 \, i = 0.415\pm0.028 \, {\rm M}_{\odot}$. This would make the minimum mass of the observed star (i.e.\ if $i=90^{\circ}$) $M_{2,{\rm min}}=1.0 \, {\rm M}_{\odot}$.

We can draw several conclusions about the nature of the primary. Fitting a single Gaussian to the H$\alpha$ emission line reveals a FWHM $\sim 450  {\rm \, km \, s^{-1}}$. LMXBs, particularly those containing BHs and with high accretion rates, typically exhibit FWHM $\gtrsim 1000  {\rm \, km \, s^{-1}}$, whereas CVs are seen to exhibit FWHM $\gtrsim 300 {\rm \, km \, s^{-1}}$ \citep[see e.g.][]{Casares-2015}, so this suggests a white dwarf accretor.

The {\em XMM-Newton} spectrum 
shows evidence of emission features at $\sim1 {\rm \, keV}$, consistent with the Fe {\sc xxiii/xxiv} L-shell transition complex \citep[][in prep.]{Bahramian-2020-prep}. Such features are typical of CVs \citep[e.g.][]{Ramsay-2001,Mukai-2003}, but have occasionally been seen in NS-LMXBs \citep[e.g.][]{vandenEijnden-2018c}, 
so this evidence alone does not 
rule out a NS primary. 

We can examine the companion a number of ways in order to draw more solid conclusions about its evolutionary state. Using an approximation for calculating the stellar effective temperature from $(B-V)_{\rm int}$ \citep{Sekiguchi-2000,Ballesteros-2012} we find $T_{\rm eff}\approx4200 {\rm \, K}$. (We calculate $T_{\rm eff}\approx4500 {\rm \, K}$ if we assume a giant donor and utilise the relationship between EW[$^{12}$CO] and $T_{\rm eff}$ \citep{Ramirez-1997} and $T_{\rm eff}\approx3600 {\rm \, K}$ assuming a dwarf \citep{Ali-1995}. {\em Gaia} DR2 reports $T_{\rm eff}=4019^{+188}_{-150} {\rm \, K}$.) 
The donor would be a K3 
giant \citep[e.g.][]{vanBelle-1999}, therefore a slightly later type subgiant. 
To convert $M_V$ to $L_{\rm bol}$, we assume an extreme case, for example in the case of a giant star of a similar spectral type, to calculate an upper limit to $L_{\rm bol}$. Assuming BC$=-1.02$ \citep[A K5 giant;][]{Cox-2000} we find $L_{\rm bol}\lesssim29\,{\rm L}_{\odot}$. This places the companion of 3XMM\,J174417.2$-$293944 redward of the post-main-sequence evolutionary tracks of stars no more massive than $2.5 \, {\rm M}_{\odot}$ \citep{Schaller-1992}, which we adopt as a realistic upper limit to $M_2$. This means that $M_1<1.04 \, {\rm M}_{\odot}$ -- the primary is almost certainly a WD. The primary's maximum mass also allows us to derive a binary inclination $i>47^{\circ}$.

The upper limit to $L_{\rm bol}$ combined with the derived $T_{\rm eff}$ allows us to estimate the radius of the companion to be $R_{\star}\lesssim10 \, R_{\odot}$. 
This suggests a subgiant, 
as giants of similar temperature are typically twice as large in radius \citep[e.g.][]{vanBelle-1999,Alonso-2000}. We use Kepler's third law and the derived range of  masses for the primary to calculate the binary separation $a\approx19.8-26.9 \, R_{\odot}$. Following \citet{Eggleton-1983} the size of the Roche lobe of the secondary is 
$R_{\rm RL,2}\approx9.0-12.3 \, R_{\odot}$. The $R_{\star}\lesssim10 \, R_{\odot}$ secondary is therefore consistent with filling, or nearly filling, its Roche lobe.

We can estimate the mass-loss rate of the stellar companion 
to see 
whether the source can be powered by accretion of a stellar wind. 
Using Reimers' Law \citep{Reimers-1977} and choosing a reasonable range of companion masses, luminosities and radii (see above), we estimate the mass-loss rate from the late-type companion to be $\dot{M}_2\sim10^{-11}-10^{-10}$ ${\rm M}_{\odot}$ y$^{-1}$ \citep[see also][]{Willems-2003}. Using the standard Bondi-Hoyle-Littleton formalism \citep{Hoyle-1941,Bondi-1944} to estimate the average mass accretion rate onto the compact object we find $\dot{M}_{\rm acc}\sim10^{-13}$ ${\rm M}_{\odot}$ y$^{-1}$ using a median $\dot{M}_2=5\times10^{-11}$ ${\rm M}_{\odot}$ y$^{-1}$. This translates to an inferred bolometric luminosity $L_{\rm bol}\sim 6\times10^{29}-3\times10^{30} {\rm \, erg \, s^{-1}}$ (assuming the known range of masses for the WD primary and the \citealt{Nauenberg-1972} WD mass-radius relation), well below the measured X-ray luminosity range \citep[0.3--$10 {\rm \, keV}$ $L_X\sim5\times10^{31-33} {\rm \, erg \, s^{-1}}$, averaging $L_X\sim10^{32} {\rm \, erg \, s^{-1}}$;][in prep.]{Bahramian-2020-prep}, indicating that spherical accretion of a stellar wind is not responsible for the observed luminosity of 3XMM\,J174417.2$-$293944.

The classic picture of a CV involves a WD accreting matter via Roche-lobe overflow (RLO) from a (less massive) main sequence companion, with typical $P_{\rm orb}\lesssim6$h. However, if 3XMM\,J174417.2$-$293944 is a RLO system, with $P_{\rm orb} = 8.7092\pm0.0048 {\rm \, d}$ and $q>1$, it would have an unusually long period, the longest for any known CV \citep{Ritter-2003}. 
Several low-mass X-ray binaries with NSs in this period range are known, where mass transfer is driven by an evolved companion losing mass on the nuclear timescale \citep{Webbink-1983}, producing mass transfer rates $\sim10^{-8} \, {\rm M}_{\odot}  {\rm \, y^{-1}}$ \citep{Podsiadlowski-2002}. We note that this would give an extremely high accretion luminosity of $L\sim6\times10^{34}-3\times10^{35} {\rm \, erg \, s^{-1}}$, which would likely produce an optically thick layer around the CV radiating most of this luminosity in the UV, which is not seen from this source \citep[][in prep.]{Rivera-Sandoval-2020-prep}. Even more problematic for this interpretation is that the high mass ratio ($q>1$, i.e.\ $M_2>M_1$) of this object would lead to rapid, thermal timescale mass transfer \citep{Ivanova-2004}, with higher mass transfer rates up to $10^{-6} \, {\rm M}_{\odot}  {\rm \, y^{-1}}$ for a few $10^6$ years, a large optically thick envelope around the WD, and possibly lead to a Type Ia supernova explosion \citep{Han-2004}. The high mass transfer rate in this scenario would lead to a variety of observable effects \citep[supersoft X-rays, strong broad optical emission lines; see e.g.][]{Southwell-1996} that do not match our observations.

Alternatively, the subgiant is close to filling its Roche lobe, and its stellar wind is gravitationally focused in the direction of the WD \citep{Friend-1982}. In a focused wind case, the mass-accretion rate on to the compact object can be up to 5--20 per cent of the total wind mass-loss rate from the donor, much more efficient than the Bondi-Hoyle-Littleton approximation \citep{deVal-Borro-2017}. Again adopting the median $\dot{M}_2=5\times10^{-11} \, {\rm M}_{\odot} {\, \rm y^{-1}}$ we find in the focused wind case an inferred $L_{\rm bol}\sim9\times10^{30}-2\times10^{32} {\rm \, erg \, s^{-1}}$. A focused wind could therefore account for the observed X-ray properties of 3XMM\,J174417.2$-$293944. Focused wind accreting binaries can be seen as an evolutionary stage between symbiotic binaries and RLO systems \citep{Friend-1982}. We find the focused-wind scenario the most physically plausible to explain 3XMM\,J174417.2$-$293944.

\section{Conclusions}
We 
give 
results of optical/NIR follow-up of X-ray sources detected in the first year of the Swift Bulge Survey: 

$\bullet$ IGR\,J17445$-$2747 is a NS binary (known from X-ray bursts) with a giant companion. This is likely to be a symbiotic X-ray binary (i.e.\ the NS accretes from the giant's wind), and the first symbiotic X-ray binary showing X-ray bursts, indicating a relatively low $B$ field at birth, since symbiotic X-ray binaries have short lifetimes.

$\bullet$ An unidentified source, Swift\,J175233.9$-$290952. The near-IR spectrum of the brightest star within the error circle (VVV\,J175233.93$-$290947.66) shows no evidence for accretion, suggesting that the true optical counterpart is relatively faint, and (considering the relation between outburst optical brightness and orbital period) may have a relatively short orbital period. 

$\bullet$ 1SXPS\,J174215.0$-$291453 is a nearby binary containing at least one chromospherically active M dwarf.

$\bullet$ 3XMM\,J174417.2$-$293944 is a WD binary with a subgiant companion in an 8.7 d orbital period. The companion is more massive than the primary ($q=2.4\pm0.1$). Neither standard Bondi-Hoyle-Littleton accretion from a wind, nor Roche-lobe overflow, account for the physical picture. We argue that the X-ray emission may be driven by a focused wind; in which case this would be the first known system of a WD accreting from its companion's focused wind.

The sources detected in the first year of the SBS that we have obtained significant follow-up observations of all have drastically different natures. It is clear that faint X-ray sources cannot be attributed to one mechanism. However,  our in-depth studies of IGR\,J17445$-$2747 and 3XMM\,J174417.2$-$293944 
indicate that 
binaries with giant companions may contribute significantly to the VFXT population. The SBS commenced its second year of operations in April 2019, and we are proceeding with optical/NIR follow-up of transient X-ray sources to extend our understanding of these sources. 

\label{sec:Conclusions}

\section*{Acknowledgements}
The authors thank the anonymous referee for useful comments that helped improve the manuscript, and N. Ivanova, K. Van and R. Hatfull for useful discussions on 3XMM\,J174417.2$-$293944. AWS thanks R. Plotkin for helpful comments on revising the manuscript.

COH \& GRS acknowledge NSERC Discovery Grants RGPIN-2016-04602 and  RGPIN-2016-06569 respectively, and COH also a Discovery Accelerator Supplement. TJM and LERS thank NASA for support under grant 80NSSC17K0334. JS acknowledges support from a Packard Fellowship. ND is supported by a Vidi grant from the Netherlands Organisation for Scientific Research (NWO).

Based on observations obtained at the Gemini Observatory (processed using the Gemini {\sc iraf} package), which is operated by the Association of Universities for Research in Astronomy, Inc., under a cooperative agreement with the NSF on behalf of the Gemini partnership: the National Science Foundation (United States), National Research Council (Canada), CONICYT (Chile), Ministerio de Ciencia, Tecnolog\'{i}a e Innovaci\'{o}n Productiva (Argentina), Minist\'{e}rio da Ci\^{e}ncia, Tecnologia e Inova\c{c}\~{a}o (Brazil), and Korea Astronomy and Space Science Institute (Republic of Korea). The authors also wish to recognize and acknowledge the very significant cultural role and reverence that the summit of Maunakea has always had within the indigenous Hawaiian community. We are most fortunate to have the opportunity to conduct observations from this mountain. Based on observations collected at the European Organisation for Astronomical Research in the Southern Hemisphere under ESO programmes 099.D-0826(C), 099.D-0826(D) and 099.A-9025(A). Part of the funding for GROND (both hardware as well as personnel) was generously granted from the Leibniz-Prize to Prof. G. Hasinger (DFG grant HA 1850/28-1). Based on observations obtained at the Southern Astrophysical Research (SOAR) telescope, which is a joint project of the Minist\'{e}rio da Ci\^{e}ncia, Tecnologia, Inova\c{c}\~{a}os e Comunica\c{c}\~{a}oes (MCTIC) do Brasil, the U.S. National Optical Astronomy Observatory (NOAO), the University of North Carolina at Chapel Hill (UNC), and Michigan State University (MSU). This work has made use of data from the European Space Agency (ESA) mission {\it Gaia} (\url{https://www.cosmos.esa.int/gaia}), processed by the {\it Gaia} Data Processing and Analysis Consortium (DPAC; \url{https://www.cosmos.esa.int/web/gaia/dpac/consortium}). Funding for the DPAC has been provided by national institutions, in particular the institutions participating in the {\it Gaia} Multilateral Agreement. This research made use of {\sc astropy} (https://www.astropy.org),
a community-developed core {\sc python} package for Astronomy \citep{Astropy-2013,Astropy-2018}.




\bibliographystyle{mnras}
\bibliography{SBS_OIR_paper.arxiv.bib} 



\appendix 
\section{Follow-up targets}
\label{app:sources}
{\bf IGR\,J17445$-$2747} was discovered by the {\em International Gamma-Ray Astrophysics Laboratory} \citep[{\em INTEGRAL};][]{Winkler-2003} and catalogued as an unidentified, variable source \citep{Bird-2006,Bird-2007,Bird-2010,Krivonos-2007,Krivonos-2010}. {\em Swift}/XRT detected activity from IGR\,J17445$-$2747 during the first epoch of the SBS on 2017 April 13 \citep{Heinke-2017}. Three days prior to the {\em Swift}/XRT detection, {\em INTEGRAL} had detected a thermonuclear burst from the source, identifying the compact object as a NS \citep{Mereminskiy-2017}. Analysis of the X-ray burst by \citet{Mereminskiy-2017} implied a minimum distance to the source $d\gtrsim5$ kpc. A short ($\sim1$ ks) \cha\ X-ray Observatory \citep{Weisskopf-2000} observation of the source in quiescence allowed precise localisation \citep{Chakrabarty-2017}. The X-ray position matches a bright ($K_s=9.65\pm0.06$) source in multiple NIR catalogues, including the 
2MASS catalogue  (2MASS\,J17443041$-$2746004) and the mid-IR Galactic Legacy Infrared Mid-Plane Survey Extraordinaire \citep[GLIMPSE;][]{Benjamin-2003}. It also matches a faint, red optical source in the {\em Panoramic Survey Telescope and Rapid Response System} ({\em Pan-STARRS}) 
survey \citep[$i=19.38\pm0.01$;][]{Chambers-2016}, but is not detected in filters bluer than $i$. 
gives the distance 
of the optical counterpart ({\em Gaia} DR2 4060626256817246720) 
as $1.1<d<7.6$ kpc \citep{Bailer-Jones-2018}.

{\bf Swift\,J175233.9$-$290952} was discovered on 2017 May 4 in epoch 3 of the SBS. Follow-up with \cha\ on 2017 May 25 provided an accurate position \citep{Maccarone-2017b}. 
The X-ray position is consistent with a NIR source in the {\em Visible and Infrared Survey Telescope for Astronomy} ({\em VISTA}) Variable in the Via Lactea Survey \citep[VVV;][]{Minniti-2010} catalogue, VVV\,J175233.93$-$290947.66. There are no 
optical or mid-IR catalogued sources here. Follow-up radio observations with the {\em Karl G. Jansky Very Large Array} ({\em VLA}) on 2017 May 13 did not detect a counterpart, with a $3\sigma$ upper limit to the luminosity $L_{10 \rm \, GHz}\lesssim1.8\times10^{26} \ (d/8{\rm \,  kpc})^2 {\rm \, erg \, s^{-1}}$ \citep{TetarenkoA-2017}.

On 2017 May 19 (epoch 4 of the SBS) we detected activity from a source at the position of the unidentified X-ray source {\bf 1SXPS\,J174215.0$-$291453} \citep{Maccarone-2017a}, and the {\em XMM-Newton} source 3XMM\,J174214.9$-$291459 
\citep{Rosen-2016}.  The XMM-Newton position is consistent with a bright source in optical ({\em Gaia} DR2 4057126472597377152) and NIR (2MASS\,J17421498$-$2914590) catalogues, 
at a distance of $77.6<d<78.2$ pc \citep{Bailer-Jones-2018}.

On 2017 September 21 (epoch 13 of the SBS) we detected an X-ray source at the position of the {\em XMM-Newton} source {\bf 3XMM\,J174417.2$-$293944} \citep[= CXO\,J174417.2$-$293943 from the {\em Chandra} source catalogue;][in prep.]{Evans-2010,Bahramian-2020-prep}. The X-ray position is consistent with a bright source in optical ({\em Gaia} DR2 4057051396569058432) and NIR (2MASS\,17441724$-$2939444) catalogues, at a {\em Gaia} distance of 
$929<d<1006$ pc \citet{Bailer-Jones-2018}.



\bsp	
\label{lastpage}
\end{document}